\documentclass[twocolumn]{aastex63}
\usepackage{graphicx}
\usepackage{enumerate}
\usepackage{amssymb, amsmath, amsfonts}
\usepackage{gensymb}
\usepackage{mathtools}
\usepackage{natbib}
\usepackage{color}
\usepackage{hyperref}
\usepackage{ulem}
\usepackage{soul}

\newcommand{\nRVstandards}{374\xspace}
\newcommand{\changes}[1]{#1}

\usepackage{xspace}

\newcommand{\hipparcos}{{\it Hipparcos}\xspace}
\newcommand{\gaia}{{\it Gaia}\xspace}

\defcitealias{Brandt_2018}{B18}

\begin{document}

\title{The Hipparcos-Gaia Catalog of Accelerations: \gaia EDR3 Edition}
\author[0000-0003-2630-8073]{Timothy D.~Brandt}
\affiliation{Department of Physics, University of California, Santa Barbara, Santa Barbara, CA 93106, USA}

\begin{abstract}
We present a cross-calibration of \hipparcos and \gaia EDR3 intended to identify astrometrically accelerating stars and to fit orbits to stars with faint, massive companions.  The resulting catalog, the EDR3 edition of the \hipparcos-\gaia Catalog of Accelerations (HGCA), provides three proper motions with calibrated uncertainties on the EDR3 reference frame: the \hipparcos proper motion, the \gaia EDR3 proper motion, and the long-term proper motion given by the difference in position between \hipparcos and \gaia EDR3.  Our approach is similar to that for the \gaia DR2 edition of the HGCA, but offers a factor of $\sim$3 improvement in precision thanks to the longer time baseline and improved data processing of \gaia EDR3.  We again find that a 60/40 mixture of the two \hipparcos reductions outperforms either reduction individually, and we find strong evidence for locally variable frame rotations between all pairs of proper motion measurements.  The substantial global frame rotation seen in DR2 proper motions has been removed in EDR3.  We also correct for color- and magnitude-dependent frame rotations at a level of up to $\sim$50~$\mu$as\,yr$^{-1}$ in \gaia EDR3.  We calibrate the \gaia EDR3 uncertainties using a sample of radial velocity standard stars without binary companions; we find an error inflation factor (a ratio of total to formal uncertainty) of 1.37.  This is substantially lower than the position dependent factor of $\sim$1.7 found for \gaia DR2 and reflects the improved data processing in EDR3.  While the catalog should be used with caution, its proper motion residuals provide a powerful tool to measure the masses and orbits of faint, massive companions to nearby stars.
\end{abstract}


\section{Introduction} \label{sec:intro}

The Early Third Data Release of the \gaia astrometry mission \citep[EDR3,][]{Gaia_EDR3} has now released position, parallax, and proper motion measurements for more than 1 billion stars \citep{Lindegren+Klioner+Hernandez+etal_2020}.  EDR3 represents a substantial improvement over the Second Data Release \citep[DR2,][]{Gaia_General_2018, Gaia_Astrometry_2018}, with proper motions improving by a magnitude-dependent factor of two to four.  \gaia EDR3 has already improved distance estimates to anchor the cosmic distance ladder \citep{Riess+Casertano+Yuan+etal_2020,Soltis+Casertano+Riess_2020} and detected proper motions of nearby dwarf galaxies \citep{McConnachie+Venn_2020}.  It has also, for the first time, provided a direct detection of the Solar acceleration in the Galactic potential \citep{Gaia_SSAccel}.

\gaia presently only performs single-star fits (generally with five astrometric parameters); accelerating and non-single stars will be treated in future data releases.\footnote{\url{https://www.cosmos.esa.int/web/gaia/release}}  However, for bright stars, the \hipparcos mission thirty years ago provides a precise and independently measured position.  
This fact was used to construct the first \gaia proper motion catalog, the {\it Tycho}-\gaia Astrometric Solution \citep[TGAS,][]{Michalik+Lindegren+Hobbs_2015,TGAS_Astrometry_2016}.  

The long-term proper motion given by the position difference between the \hipparcos and \gaia epochs, similar to TGAS, provides a precise measurement that may be compared to \gaia's proper motions.  This permits an external calibration of \gaia, and allows the catalog to be used to find and study astrometrically accelerating systems even before the release of higher-order astrometric fits by the \gaia team itself.  As released, however, \gaia DR2's proper motions are not suitable for finding accelerating stars or for fitting orbits.  A cross-calibration is required first to enforce a common reference frame and to ensure Gaussian residuals with the appropriate variance \citep[][hereafter B18]{Brandt_2018}.

\gaia EDR3 provides positions and proper motions based on 34 months of data rather than the 22 months used for DR2.  The additional data, combined with better control of systematics, result in formal proper motion uncertainties nearly a factor of 4 lower than DR2 for bright stars.  The \gaia Collaboration has performed extensive verification of the EDR3 catalog, but this has focused on stars fainter than $G \approx 10$ \citep{Lindegren+Bastian+Biermann+etal_2020,Fabricius+Luri+Arenou+etal_2020}.  The published verification analyses use a combination of binary stars, cluster measurements, and measurements of quasars and stars in the Large Magellanic Cloud.  In these cases, the parallax and/or proper motions are known, or can be compared between stars known to have the same values to well within \gaia's uncertainty.

The goal of this paper is to cross-calibrate  \hipparcos specifically to \gaia EDR3, and to extend the cross-calibration and catalog verification of \gaia EDR3 to the brightest stars present in EDR3 ($G \approx 4$).  \gaia EDR3 represents a substantial increase in precision relative to DR2, especially in its proper motion measurements.  With an updated cross-calibration, this improved precision will translate directly into better sensitivity to astrometric acceleration \citep{Kervella+Arenou+Mignard+etal_2019,Fontanive+Muzic+Bonavita+Biller_2019}, and better masses and orbits of faint companions to nearby stars \citep[e.g.][]{Brandt+Dupuy+Bowler_2019,Brandt+Dupuy+Bowler+etal_2020,Brandt+Brandt+Dupuy+etal_2020,Maire+Molaverdikhani+Desidera+etal_2020,DeRosa+Dawson+Nielsen_2020,Xuan+Wyatt_2020,Bowler+Cochran+Endl+etal_2020}.

We structure the paper as follows.  Section \ref{sec:xmatch} reviews our cross-match between \hipparcos and \gaia EDR3.  Section \ref{sec:dr2_edr3_compare} summarizes the cross-calibration results of \citetalias{Brandt_2018} with \gaia DR2, and describes how the improvements of EDR3 affect our approach.  Section \ref{sec:hipparcos} describes our construction of \hipparcos astrometric parameters and reference epochs from the two \hipparcos reductions.  Section \ref{sec:refframe} describes our local cross-calibration of the reference frames.  In Section \ref{sec:gaia_uncertainties}, we construct a reference sample and derive a spatially uniform inflation factor for the \gaia EDR3 uncertainties.  Section \ref{sec:results} summarizes the results of the cross-calibration and shows the final improvements relative to \gaia DR2.  We describe the construction and structure of the catalog in Section \ref{sec:construction_structure}.  We conclude with Section \ref{sec:conclusions}.

\section{Cross-Matching \hipparcos and \gaia EDR3} \label{sec:xmatch}

The first step in a cross-calibration of \hipparcos and \gaia EDR3 is a cross-match of sources in the two catalogs.  The \gaia archive does provide such a cross-match, but it is very incomplete.  Out of nearly 118,000 \hipparcos stars the cross-matches in the archive provide matches for just 100,000.  This reflects quality checks that we wish to relax: we want a more complete cross-match that keeps stars that accelerate between the catalogs.  

\citetalias{Brandt_2018} cross-match \hipparcos and \gaia based on a lenient positional criterion and subsequently matching colors and magnitudes.  We take a simpler approach here: we propagate the \gaia EDR3 proper motions to the \hipparcos catalog epoch of 1991.25 and search for positional matches in \hipparcos within $1^{\prime\prime}$ using the {\tt conesearch} function on the \gaia archive.  We restrict the search to stars brighter than $G=13.7$ to limit the computational demand.  

Our $1^{\prime\prime}$ cross-match returns 116,224 results.  A few hundred \hipparcos stars have multiple matches in \gaia EDR3.  For these stars we choose the brightest \gaia EDR3 source.  This leaves 115,346 matches, with one \gaia star for each \hipparcos star.  Several thousand \hipparcos stars, including many extremely bright stars with $G \lesssim 3$, do not have matches with five-parameter solutions in \gaia EDR3.  These stars saturate \gaia too heavily to recover reliable astrometry even with the most aggressive gating \citep{Lindegren+Klioner+Hernandez+etal_2020}.  We take the cross-matched sample of 115,346 stars as the starting point for the rest of our analysis.  

\section{Early DR3 vs.~DR2} \label{sec:dr2_edr3_compare}

In this section we summarize the main results from cross-calibrating \gaia DR2 to \hipparcos.  We then review the differences between DR2 and EDR3 that most significantly impact a cross-calibration.  

\begin{figure*}
\begin{center}
    \includegraphics[height=0.37\linewidth]{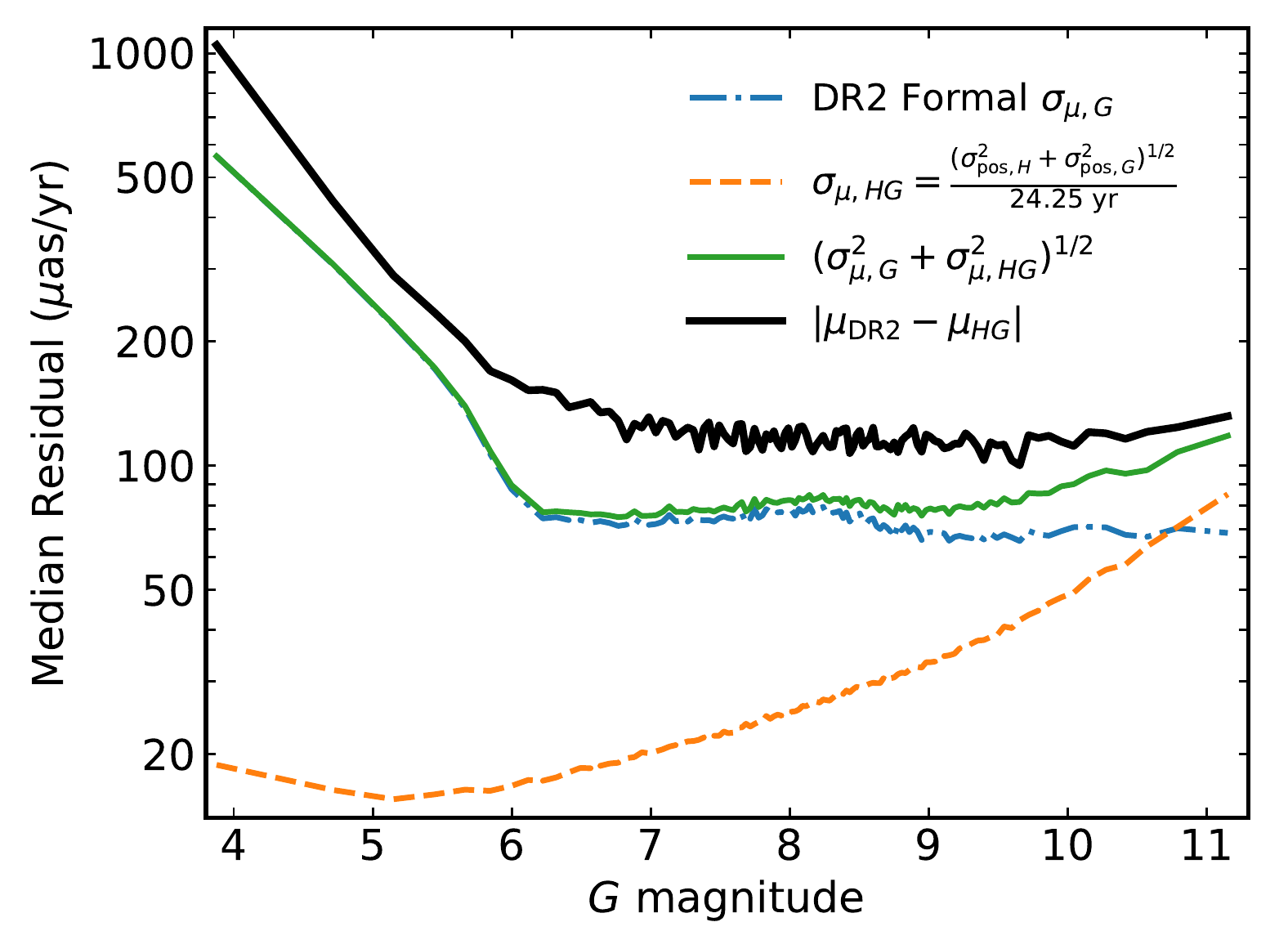} \hspace{-0.65 truein}
    \includegraphics[height=0.37\linewidth]{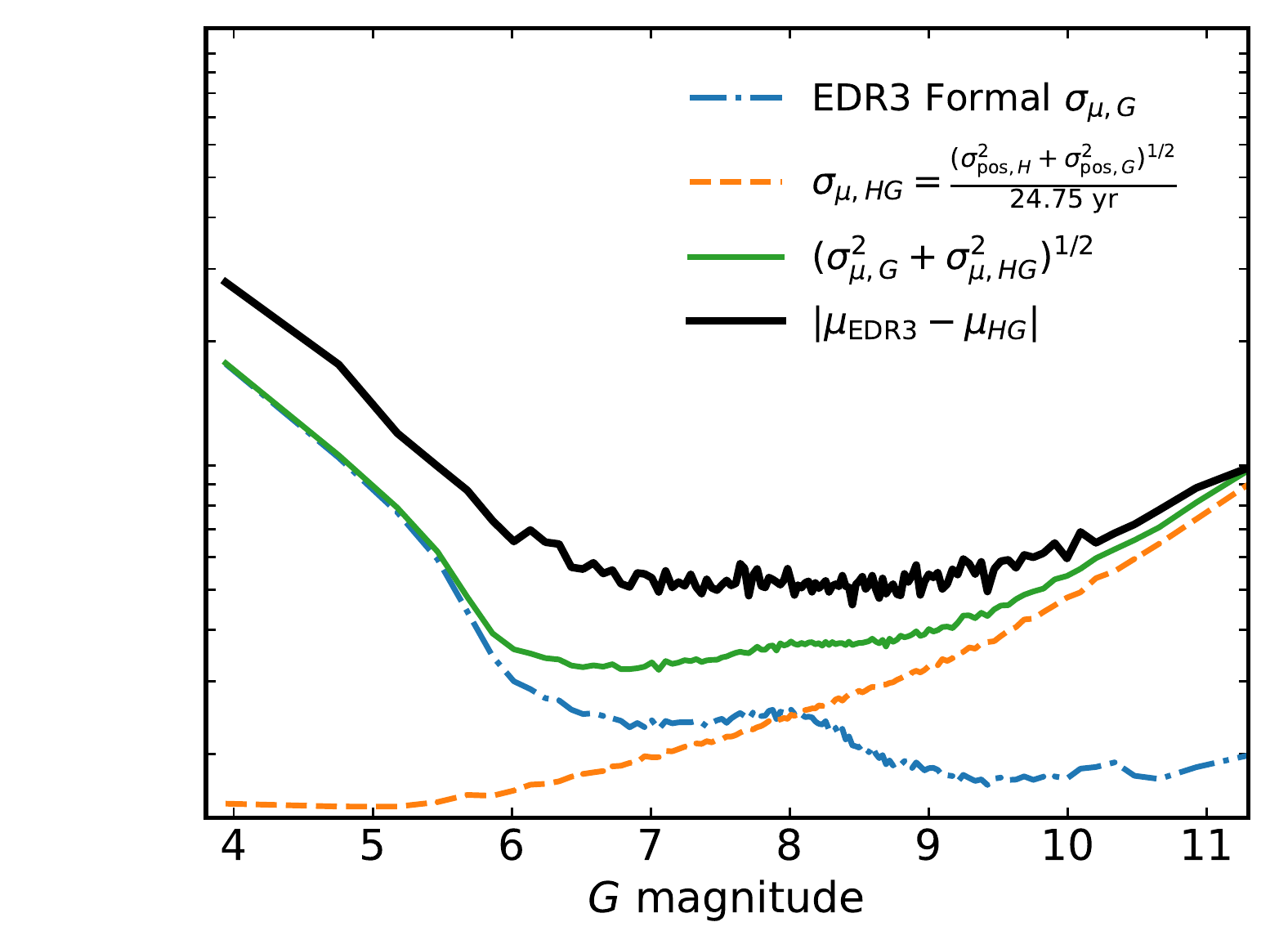}
\end{center}
\vskip -0.1 truein
    \caption{Comparison of \gaia DR2 and EDR3 precisions.  Left: the median uncertainties on DR2 proper motions (blue dot-dashed line), on long-term proper motions (orange dashed line), and their quadrature sum (green solid line) compared the median absolute difference between the DR2 and long-term proper motions (black line).  The green curve represents the formal sensitivity of \hipparcos+\gaia DR2 to astrometric acceleration. 
    Right: the same uncertainties and residuals but for \gaia EDR3 rather than DR2.  The formal uncertainties on proper motion have fallen by a magnitude-dependent factor of $\sim$3; the median astrometric residuals have also fallen accordingly.  While the long-term proper motion was the most precise measurement for nearly all \hipparcos stars in DR2, the \gaia EDR3 proper motions are now the most precise measurements for about half of \hipparcos stars.  }
    \label{fig:errors_mag}
\end{figure*}

\subsection{The \gaia DR2 HGCA}

The DR2 version of the HGCA \citepalias{Brandt_2018} represents a cross-calibration of three proper motions: the \hipparcos proper motions, the \gaia DR2 proper motions, and the position difference between \hipparcos and \gaia divided by the difference in epoch between the two surveys.  This long-term proper motion is the most precise measurement for nearly 98\% of stars after calibrating the \gaia DR2 uncertainties.  This permits a clean separation of \hipparcos and \gaia proper motions: both can separately be calibrated to the more precise long-term proper motions.  The uncertainties and systematics in \hipparcos and \gaia DR2 are, as a result, relatively straightforward to disentangle.

\citetalias{Brandt_2018} find, with $\sim$150$\sigma$ significance, that a 60/40 weighted average of the \cite{ESA_1997} and \cite{vanLeeuwen_2007} \hipparcos reductions outperforms either on its own.  They also find that the \hipparcos proper motions display small-scale structure that can be removed.  Doing so improves the agreement of the \hipparcos proper motions with the long-term proper motions in a cross-validation set at high significance.  Finally, \citetalias{Brandt_2018} find that an additional 0.2~mas\,yr$^{-1}$ has to be added in quadrature to the proper motion uncertainties to achieve a statistical agreement with the long-term proper motions.  The calibrated \hipparcos proper motion uncertainties are comparable to the reported uncertainties in the 1997 catalog \citep{ESA_1997} and larger than the uncertainties in \cite{vanLeeuwen_2007}, especially for bright stars.

For \gaia DR2, \citetalias{Brandt_2018} recover the frame rotation seen by \cite{Gaia_Astrometry_2018} with modest additional small-scale variations.  They determine a spatially variable error inflation (ratio of true to formal uncertainties) averaging 1.7.  This error inflation varies spatially, primarily with ecliptic latitude and hence with the average number of observations according to \gaia's scanning pattern.  Finally, \citetalias{Brandt_2018} find that the stars in \gaia with the largest uncertainties (predominantly the very bright stars) have underestimated uncertainties even after error inflation.

\subsection{Changes with EDR3}

\gaia EDR3 has implemented a number of improvements over DR2.  The most important for the HGCA is a factor of nearly 4 improvement in proper motion uncertainties for bright stars \citep{Lindegren+Klioner+Hernandez+etal_2020,Fabricius+Luri+Arenou+etal_2020}.  A factor of $(34\,{\rm months}/22\,{\rm months})^{3/2} \approx 2$ is attributable to the longer observational baseline; the rest is due to improved data processing and control of systematics \citep{Fabricius+Luri+Arenou+etal_2020}.  \gaia EDR3 has also removed most of the frame rotation that affected the bright star reference frame of DR2, and it has corrected the DOF bug \citep{Lindegren+Klioner+Hernandez+etal_2020}.  

Figure \ref{fig:errors_mag} shows the improvements from DR2 (on the left) to EDR3 (on the right).  \changes{In this and subsequent figures, we show proper motions in right ascension and declination together.  All of our quantitative analyses include the separate uncertainties and covariance of the two components of proper motion.  From DR2 to EDR3,} the formal proper motion uncertainties have fallen by a magnitude-dependent factor of three or four, while the median residuals between the long-term proper motion and the \gaia proper motion (corrected for frame rotation in DR2's case) have fallen by a factor of two to three.  The long-term proper motions adopt the 60/40 mix of the two \hipparcos reductions found by \citetalias{Brandt_2018}.  In DR2, the long-term proper motion uncertainties were dominated by \gaia errors for very bright magnitudes ($G \lesssim 5$).  They are now dominated by \hipparcos uncertainties at all magnitudes.

In the DR2 edition of the HGCA, the median precision on the long-term proper motion was a factor of $\sim$4 better than the \gaia proper motions.  Structure and systematics in the proper motion residuals were overwhelmingly due to the rotation of \gaia's bright star reference frame.  The \gaia EDR3 and long-term proper motions are comparably precise.  Their difference is now equally sensitive to systematic rotations in the \gaia reference frame and positional systematics in the \hipparcos reference frame.  Despite the difficulty in interpreting structures and systematics in the proper motion residuals, we follow the same approach to cross-calibrate the catalogs.

\begin{figure}
    \includegraphics[width=\linewidth]{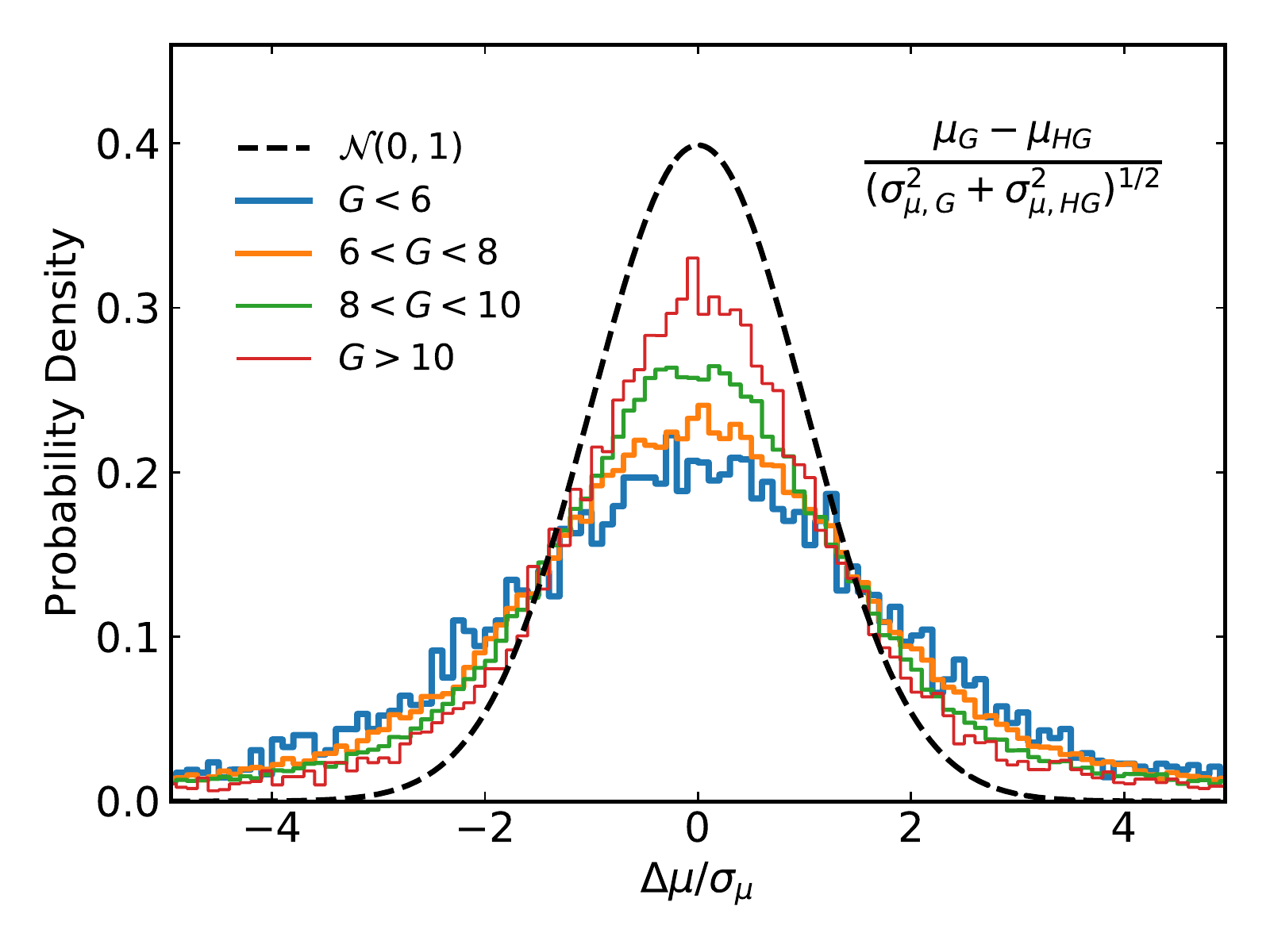}
    \caption{Residuals between the \gaia EDR3 proper motions and the long-term proper motions, normalized by their combined uncertainty.  \changes{We compute the normalized residuals separately for right ascension and declination and include both in the histograms.} The residuals are much narrower than for DR2 (c.f.~Figure 1 of \citetalias{Brandt_2018}) but still show a significant magnitude dependence.  At the brighter magnitudes where \gaia EDR3 proper motion uncertainties dominate the error budget, the distributions are substantially wider than a unit Gaussian and reflect underestimated uncertainties.}
    \label{fig:pm_residuals_raw}
\end{figure}

Figure \ref{fig:pm_residuals_raw} shows the distribution of $z$-scores: the residuals between the \gaia EDR3 and the long-term proper motions, divided by their standard errors.  \changes{In this and subsequent figures, we compute $z$-scores in right ascension and declination separately and then plot them together.  The uncertainties on both are Gaussian even when the two components of proper motion are covariant.}  The standard errors are the quadrature sums of the formal uncertainties in the EDR3 proper motions and the positional uncertainties in EDR3 and the merged \hipparcos catalog, scaled by the 24.75 year baseline between the catalogs.  The distributions deviate from a unit Gaussian.  The width of the distribution increases at bright magnitudes where \gaia EDR3 proper motions dominate the error budget.  This suggests that the EDR3 formal uncertainties underestimate the true uncertainties.  This underestimate is expected \citep{Lindegren+Lammers+Hobbs+etal_2012} and also seen by the \gaia team \citep{Lindegren+Klioner+Hernandez+etal_2020,Fabricius+Luri+Arenou+etal_2020}.  The distributions of Figure \ref{fig:pm_residuals_raw} also have tails well in excess of those of a Gaussian; these are likely from stars that are astrometrically accelerating.  

Figure \ref{fig:pm_residuals_raw} compares favorably to Figure 1 of \citetalias{Brandt_2018}, the equivalent plot for \gaia DR2.  The systematics from frame rotation are mostly gone and the formal uncertainties appear to be closer to the true uncertainties.  A cross-calibration of \hipparcos and \gaia remains necessary, but the catalogs as published are in much closer agreement than they were for \gaia DR2.

\section{Hipparcos Positions, Proper Motions, and Epochs} \label{sec:hipparcos}

The cross-calibration of \hipparcos in the HGCA \citepalias{Brandt_2018} relies on \gaia positions\changes{, proper motions,} and parallaxes.  While \changes{all} are more precise in EDR3, \gaia \changes{ astrometry was} already far more precise than the \hipparcos values for most stars in DR2.  Our approach and results to the \hipparcos proper motions are essentially the same here as in \citetalias{Brandt_2018}.  This applies especially to the linear combination of the two \hipparcos reductions used by \citetalias{Brandt_2018} and the use of \gaia parallaxes to refine the other \hipparcos astrometric parameters.  

We summarize the calibration steps for the \hipparcos catalogs here.  All of these steps closely match steps taken for the DR2 edition of the HGCA \citepalias{Brandt_2018}.

\subsection{Use of the Gaia EDR3 parallaxes}

We use the \gaia EDR3 parallax measurements and uncertainties to refine the other \hipparcos astrometric parameters in the same way as \citetalias{Brandt_2018}.  We refer to Section 4 of that paper for details.  For \gaia DR2, \citetalias{Brandt_2018} found an improvement of $\sim$1\% in the agreement between \hipparcos and \gaia proper motions after incorporating \gaia parallax measurements to refine the other \hipparcos astrometric parameters.  We do not apply an error inflation to the \gaia EDR3 parallaxes for this purpose.  \cite{El-Badry+Rix+Heintz_2021} found that the EDR3 parallax uncertainties were underestimated by $\lesssim$30\% for stars with well-behaved fits.

We find that the EDR3 parallaxes offer the same $\sim$1\% improvement to the \hipparcos proper motions as the DR2 parallaxes did.  The DR2 parallaxes were precise enough relative to the \hipparcos measurements that EDR3 provides a negligible gain.  The parallaxes also improve the \hipparcos positions: the mean absolute deviation between the long-term proper motions (whose uncertainties are dominated by \hipparcos) and the \gaia EDR3 proper motions fall by about 0.4\% in right ascension and 0.2\% in declination when incorporating \gaia parallaxes into the \hipparcos solution.  

Finally, we assess whether correcting for the parallax bias according to the prescriptions derived by \cite{Lindegren+Bastian+Biermann+etal_2020} can further improve \hipparcos astrometry.  These bias corrections are functions of source magnitude, color, and ecliptic latitude.  We find that applying these corrections to EDR3 parallaxes offers a tiny improvement in the agreement of \hipparcos proper motions with long-term proper motions.  However, it slightly degrades the agreement between long-term proper motions and \gaia EDR3 proper motions.  As a result, we do not apply any bias corrections to the EDR3 parallaxes.  The effects of applying a bias correction to the EDR3 parallaxes are at least two orders of magnitude smaller than the improvement from using the EDR3 parallaxes directly.  

Our analysis says little about the accuracy of the \cite{Lindegren+Bastian+Biermann+etal_2020} bias corrections at bright magnitudes.  The biases are much smaller than the \hipparcos parallax uncertainties, especially around $G \sim 12$ where the corrections change abruptly with magnitude \citep{Lindegren+Bastian+Biermann+etal_2020}.  Further, we can only measure the agreement between \gaia parallaxes and true parallaxes in the \hipparcos frame, which could be subject to their own biases and systematics.

\subsection{The Astrometric Reference Epoch} \label{sec:ref_epoch}

Our next step is to propagate positions to the central epochs of the \hipparcos observations (the epochs that minimize the positional uncertainties), exactly as was done by \citetalias{Brandt_2018}.  The central epochs generally differ in right ascension and declination, and they also differ from the catalog epoch of 1991.25. This step results in a smaller uncertainty on position and removes covariance between position and proper motion.  It also provides a more accurate approximation to the true epoch at which the proper motion is effectively measured.  We apply this exact same approach to \gaia EDR3 using its covariance matrix.

The perturbation to the reference epoch in right ascension is given by
\begin{equation}
    \delta t_{\alpha*} = - \frac{{\rm Cov}(\alpha*, \mu_{\alpha*})}{\sigma^2[\mu_{\alpha*}]}
    \label{eq:deltat}
\end{equation}
where ${\rm Cov}(\alpha*, \mu_{\alpha*})$ is the covariance between position and proper motion in right ascension ($\alpha* \equiv \alpha \cos \delta$) and $\sigma^2[\mu_{\alpha*}]$ is the variance in proper motion.  We compute Equation \eqref{eq:deltat} and its counterpart for declination separately and propagate the positions to $t_{\rm catalog} + \delta t_{\alpha*}$ and $t_{\rm catalog} + \delta t_{\delta}$.  We then update the covariance matrices for the astrometric parameters as described in \citetalias{Brandt_2018}.

With the astrometric reference epoch defined as above, we compute the long-term proper motions as, e.g.,
\begin{equation}
    \mu_{HG, \alpha*} = \left( \frac{\alpha_{\it Gaia} - \alpha_{\it Hip}}{t_{\it Gaia,\alpha} - t_{\it Hip, \alpha}} \right) \cos \left[ \frac{\delta_{\it Gaia} + \delta_{\it Hip}}{2} \right].
\end{equation}
Once calibrated relative to the \hipparcos and \gaia EDR3 reference frames, this serves as one of the three proper motions supplied by the catalog.

\subsection{The composite Hipparcos catalog}

\citetalias{Brandt_2018} adopted a 60/40 mixture of the two \hipparcos reductions for both position and proper motion, with an additional uncertainty of 0.2~mas\,yr$^{-1}$ added in quadrature with the \hipparcos proper motions.  The relative weights of the \cite{ESA_1997} and \cite{vanLeeuwen_2007} reductions and the additional proper motion uncertainty were driven
almost entirely by the \gaia DR2 position measurements and the \hipparcos proper motions.  While the \gaia positions have improved with EDR3, they were already so precise with DR2 that the additional precision has no effect on the main results of \citetalias{Brandt_2018}.  

We perform one more check on the \hipparcos catalog mixture.  The \gaia EDR3 proper motions are now comparable in precision to the long-term proper motions.  We therefore compute the \hipparcos weights that give the best agreement between the \gaia EDR3 proper motions and the long-term proper motions.  This is sensitive to the \hipparcos positions, but independent of the \hipparcos proper motions.  We again find that a 60/40 mix is optimal, in agreement with the optimal mix found by \citetalias{Brandt_2018} for the \hipparcos proper motions.  We adopt this 60/40 linear combination of the two \hipparcos catalogs, and an error inflation of 0.2~mas\,yr$^{-1}$ added in quadrature, from \citetalias{Brandt_2018}.

\section{A Common Reference Frame} \label{sec:refframe}

Our next step is to place the \hipparcos proper motions, the \gaia EDR3 proper motions, and the long-term proper motions on a common reference frame.  Our approach is the same as that in \citetalias{Brandt_2018}, and we refer to that paper for full details.  We provide a brief summary of the approach here before discussing our results.  

\citetalias{Brandt_2018} first fit for global rotations between the three reference frames defined by the \hipparcos proper motions, the \gaia DR2 proper motions, and the long-term proper motions, but found that locally variable rotations were statistically preferred.  We therefore begin with locally variable calibrations here.  Our approach is to divide the sky into small tiles with approximately equal numbers of stars, fit for cross-calibration parameters in each one, and then produce a smoothed map for the correction using a Gaussian process regression with a Mat\'ern covariance function.  We do not know, a priori, how much structure there is in the residuals between the three reference frames.  We therefore tile the sky at a range of resolutions in order to find the one that provides the best cross-calibration.  

We wish to use only stars that have constant proper motion to calibrate the reference frames.  We therefore discard stars for which any proper motion is inconsistent at $>10\sigma$ between \hipparcos, \gaia, and/or the long-term proper motion.  This step removes about 28,000 stars (most of which are physically accelerating), leaving just over 87,000 to match the reference frames.  Ever-finer tilings of the sky also carry a risk of overfitting, so we hold back 10\% of the stars 
as a cross-validation data set.  We also hold back our radial velocity reference stars (described in Section \ref{sec:rv_references}) to avoid biasing our eventual calibration of the \gaia EDR3 proper motion uncertainties.

Given a small patch of sky and a small number of stars in that patch, we fit for the frame rotation that maximizes the likelihood of a Gaussian mixture model
\begin{align}
    {\cal L} &= \prod_{{\rm stars}~i} \Bigg\{ \frac{g}{2\pi \sqrt{\det {\bf C}_i}} \exp\left[ -\frac{\chi_i^2}{2} \right]  \nonumber \\ 
    &\qquad\qquad + \frac{1 - g}{2\pi \sigma^2} \exp\left[ -\left( \frac{\left( \Delta \mu_{\alpha*,i}\right)^2 + \left(\Delta \mu_\delta,i \right)^2}{2\sigma^2} \right)\right] \Bigg\}
    \label{eq:gaussmix}
\end{align}
with
\begin{equation}
    \chi_i^2 = 
    \begin{bmatrix}
    \Delta \mu_{\alpha*,i} & \Delta \mu_{\delta,i}
    \end{bmatrix}
    {\bf C}_i^{-1}
    \begin{bmatrix}
    \Delta \mu_{\alpha*,i} \\
    \Delta \mu_{\delta,i}
    \end{bmatrix}.
\end{equation}
\changes{In Equation \eqref{eq:gaussmix}, $\sigma$ is the width of a distribution to capture outliers,} and $g$, the prior for a star \changes{not} to be an outlier, \changes{is} set to $g=1/2$; \changes{both $g$ and $\sigma$ are the same for all stars}.  For \hipparcos proper motions, we use the covariance matrix ${\bf C}$ given by the weighted sum of the covariance matrix from each \hipparcos reduction, with a weight of 0.6 applied to the \cite{vanLeeuwen_2007} reduction and a weight of 0.4 to the \cite{ESA_1997} reduction.\footnote{These coefficients would be $0.4^2$ and $0.6^2$ if the two \hipparcos reductions were independent.  However, they are based on the same observations.  The improved agreement with long-term proper motions is much less than the factor of $\sim \sqrt{2}$ that would be expected for truly independent data sets.}  We then add an additional 0.2~mas\,yr$^{-1}$ in quadrature along the diagonal \citepalias{Brandt_2018}.  We do not add any additional uncertainty to the positions.  In a cross-calibration, an additive uncertainty to the \hipparcos positions is fully degenerate with an additive uncertainty to the \gaia EDR3 proper motions.  We defer that analysis to Section \ref{sec:gaia_uncertainties}.

We fit for a frame rotation appropriate to all stars in each patch of sky.  The frame rotation on an axis passing through this tile is unconstrained, so we set it to zero and constrain only the other two components.  We maximize the likelihood given by Equation \eqref{eq:gaussmix}.  Finally, we use the cross-validation sample to constrain the hyperparameters of a Gaussian Process Regression and to determine the optimal tiling of the sky.

\subsection{Nonlinear Motion} 

A star moving uniformly through space will apparently accelerate when its motion is projected onto spherical coordinates.  We take each star's instantaneous position, proper motion, parallax, and radial velocity in \gaia EDR3 and convert it \changes{into} a three-dimensional position and space velocity.  For stars without radial velocities in EDR3, we use radial velocities taken from the Extended \hipparcos compilation \citep[XHIP,][]{XHIP_2012}.  About 80\% of \hipparcos stars have radial velocities from one source or the other.  We assume zero radial velocity when computing three-dimensional velocities for the remaining 20\% of stars.

We propagate three-dimensional positions backwards using each star's three-dimensional velocity.  We propagate by the time difference between \hipparcos and \gaia EDR3; this time varies from star to star and between right ascension and declination (Section \ref{sec:ref_epoch}).  We then measure the difference between the correctly propagated position $\delta x$ and the linearly propagated position $\delta x_{\rm lin} = \mu \delta t$ divided by this time.  We apply this nonlinearity correction to the difference between the long-term proper motion and the \gaia proper motion.  We apply the same correction between the \hipparcos proper motion and the long-term proper motion; this is equivalent to applying twice the correction to the difference between the \hipparcos and the \gaia proper motions.  

We apply our computed nonlinearity corrections before calibrating the reference frames between \hipparcos, \gaia EDR3, and the long-term proper motions.  The corrections are small for most stars but reach 16 mas\,yr$^{-1}$ (in declination) for Barnard's Star.  For all but fifteen stars, the nonlinearity corrections in both right ascension and declination are less than 1~mas\,yr$^{-1}$.

\subsection{The Hipparcos Proper Motions}

\begin{figure*}
\vskip -0.4 truein
    \includegraphics[width=0.5\textwidth]{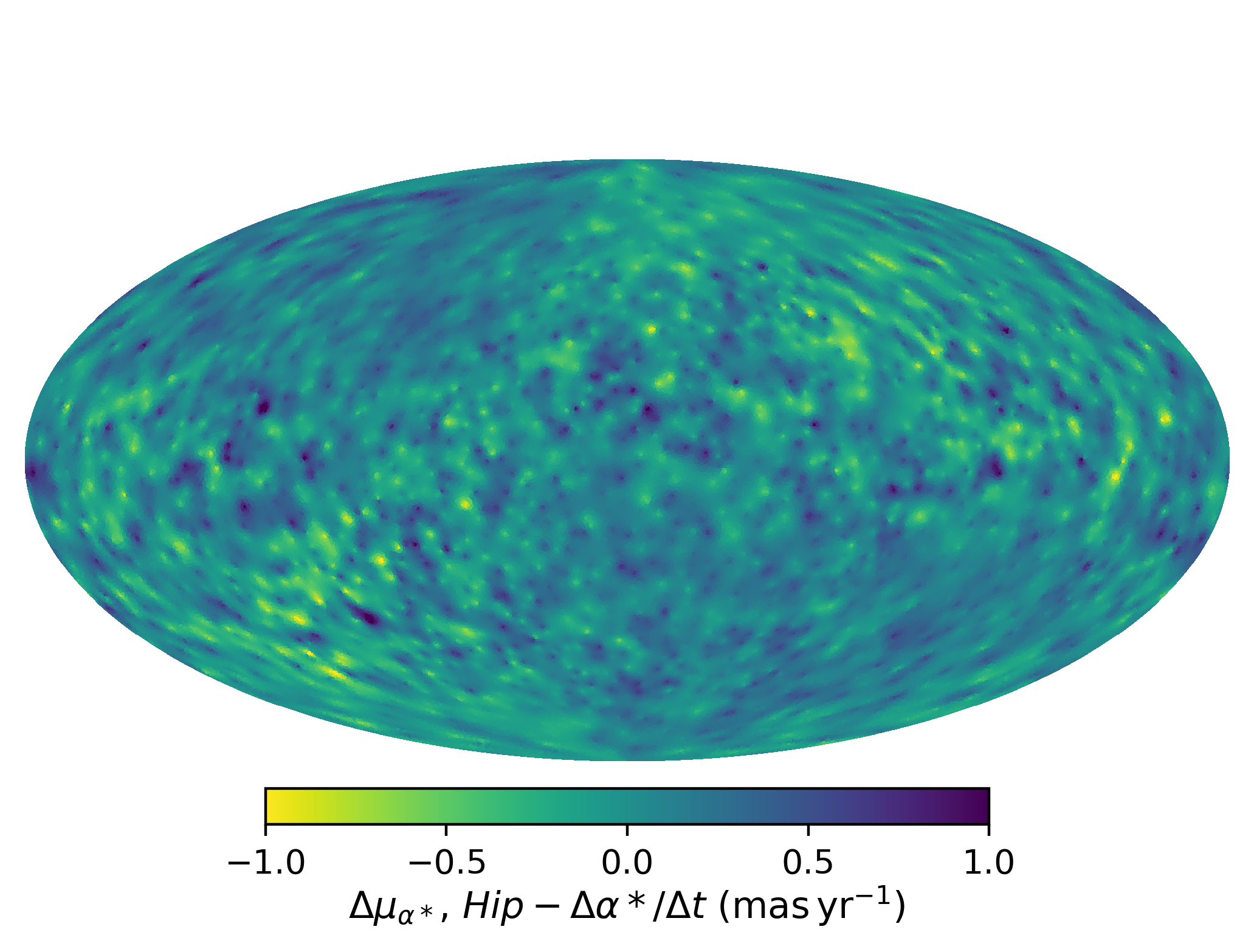}
    \includegraphics[width=0.5\textwidth]{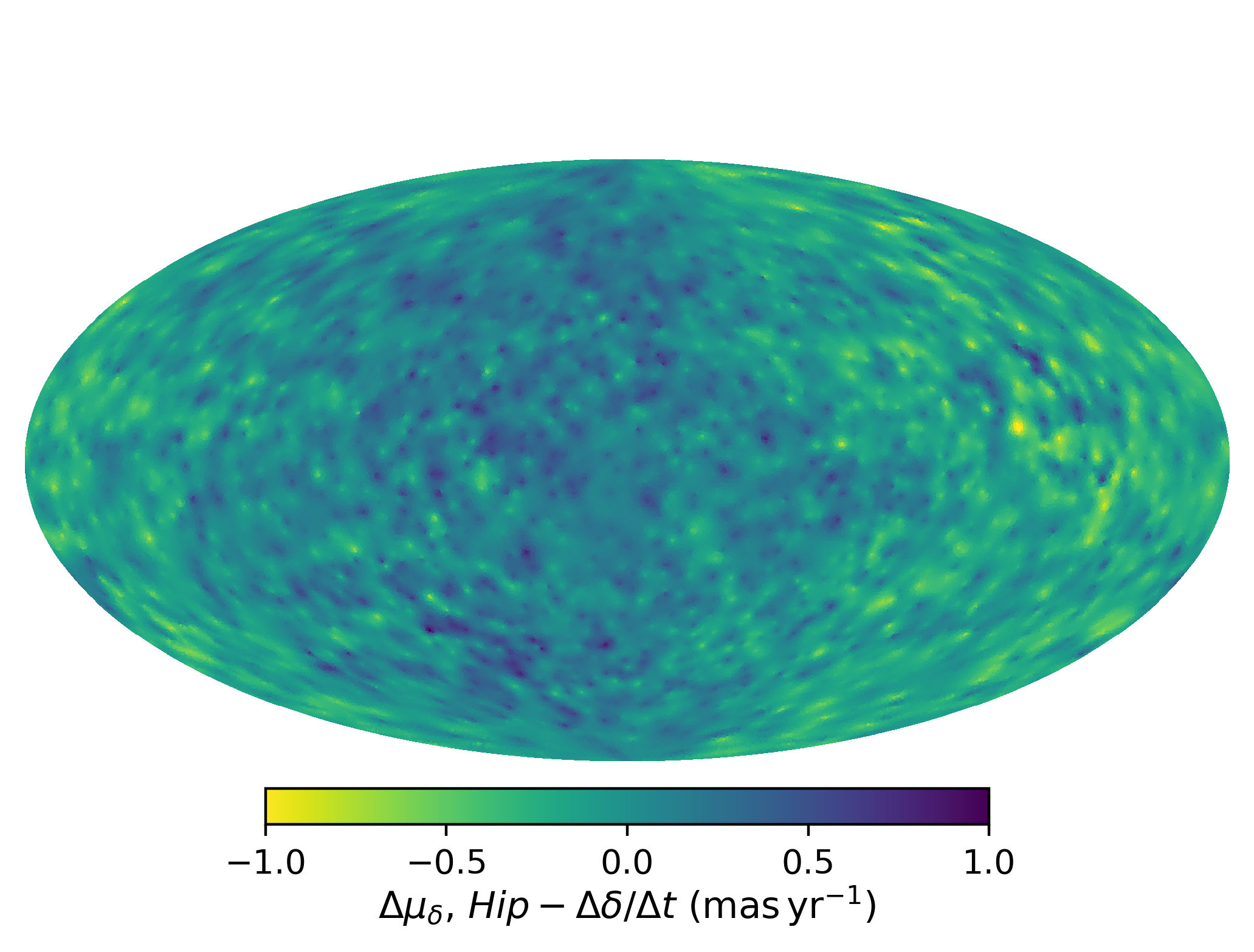}
    \caption{Local corrections to the \hipparcos proper motion relative to the long-term proper motion.  These corrections have a standard deviation of 0.2~mas\changes{\,yr$^{-1}$} in both right ascension and declination, with a few stars having corrections above 1 mas\changes{\,yr$^{-1}$}.  The corrections are very similar to those derived by \citetalias{Brandt_2018} for the DR2 HGCA, as the positions in \gaia DR2 were already far more precise than for nearly all \hipparcos stars.  The better parallaxes in EDR3 allow a slight refinement of the \hipparcos astrometry and result in slightly different corrections.  The use of a cross-validation set of stars suggest mild overfitting.  The maps show Hammer projections in equatorial coordinates with $\alpha = \delta = 0$ at the center.}
    \label{fig:hip_pmcorrect}
\end{figure*}

The \hipparcos data themselves have not changed since \gaia DR2 (apart from tiny changes in the \gaia parallax used to refine \hipparcos astrometry).  As a result, our cross-calibration between the \hipparcos proper motions and the long-term proper motions is nearly identical to the results of \citetalias{Brandt_2018}.  It does differ very slightly because we use a different set of stars to fit the reference frame and for cross-validation: \citetalias{Brandt_2018} used stars with \hipparcos IDs ending in zero, while we use stars with a \gaia EDR3 random index ending in zero.  

Figure \ref{fig:hip_pmcorrect} shows the local corrections to the \hipparcos proper motions to bring them into agreement with the long-term proper motions.  These plots are very similar to Figure 6 of \citetalias{Brandt_2018}.  Similarly as for the DR2 edition of the HGCA, these corrections to the \hipparcos proper motion frame increase the likelihood of the 8828 stars used for cross-validation by about $e^{340}$ over a constant frame rotation, and by about $e^{550}$ over no frame rotation at all.  For the 87,386 stars consistent at $10\sigma$ with constant proper motion, the correction of Figure \ref{fig:hip_pmcorrect} increases the likelihood by a factor of about $e^{5000}$ over a constant frame rotation.  This is somewhat more than the tenth power of $e^{340}$,  suggesting a modest degree of overfitting just as for the DR2 edition of the HGCA.

\subsection{The Gaia Proper Motions}

\begin{figure*}
\vskip -0.4 truein
\includegraphics[width=0.5\textwidth]{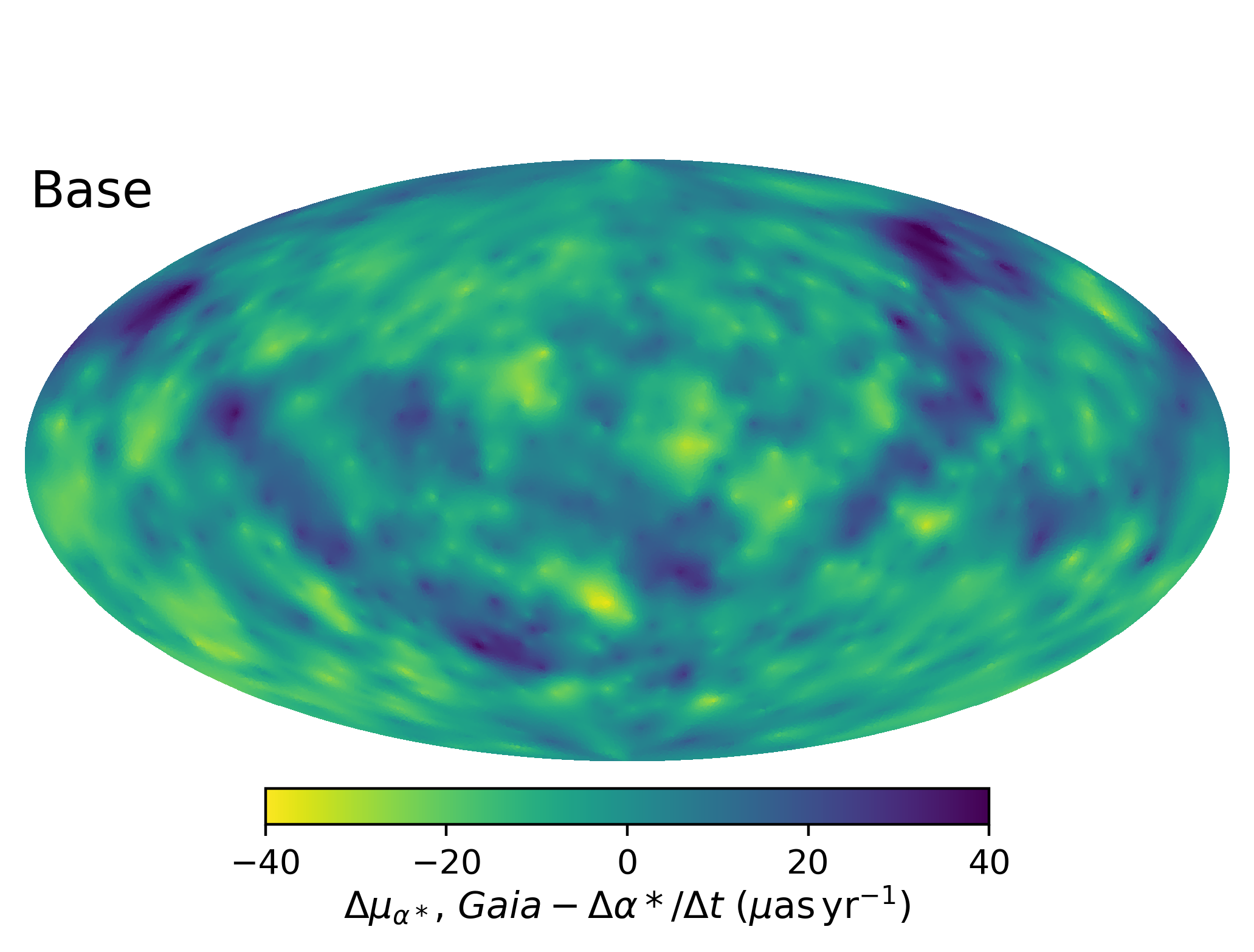}
\includegraphics[width=0.5\textwidth]{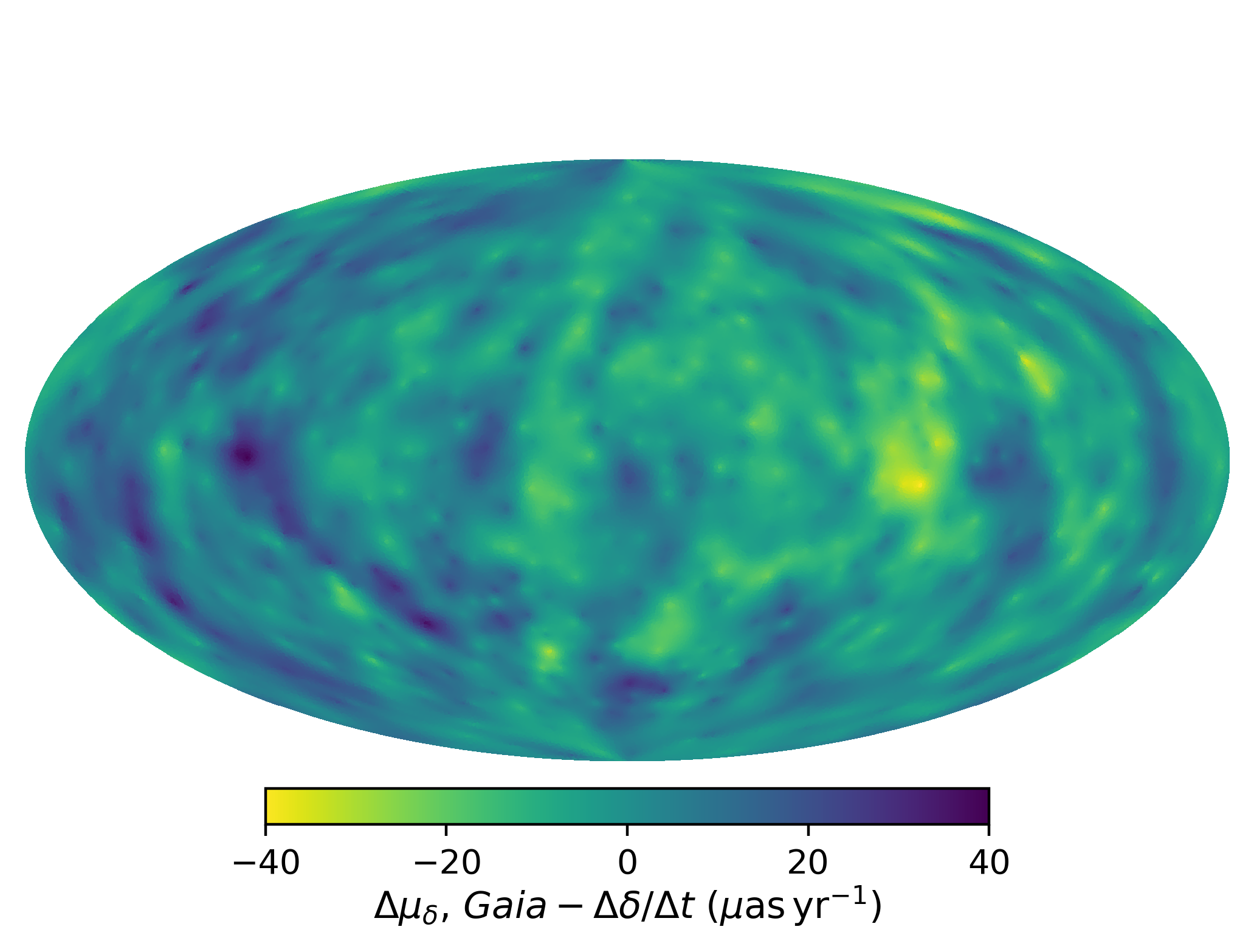}
\vskip -0.4 truein
\includegraphics[width=0.5\textwidth]{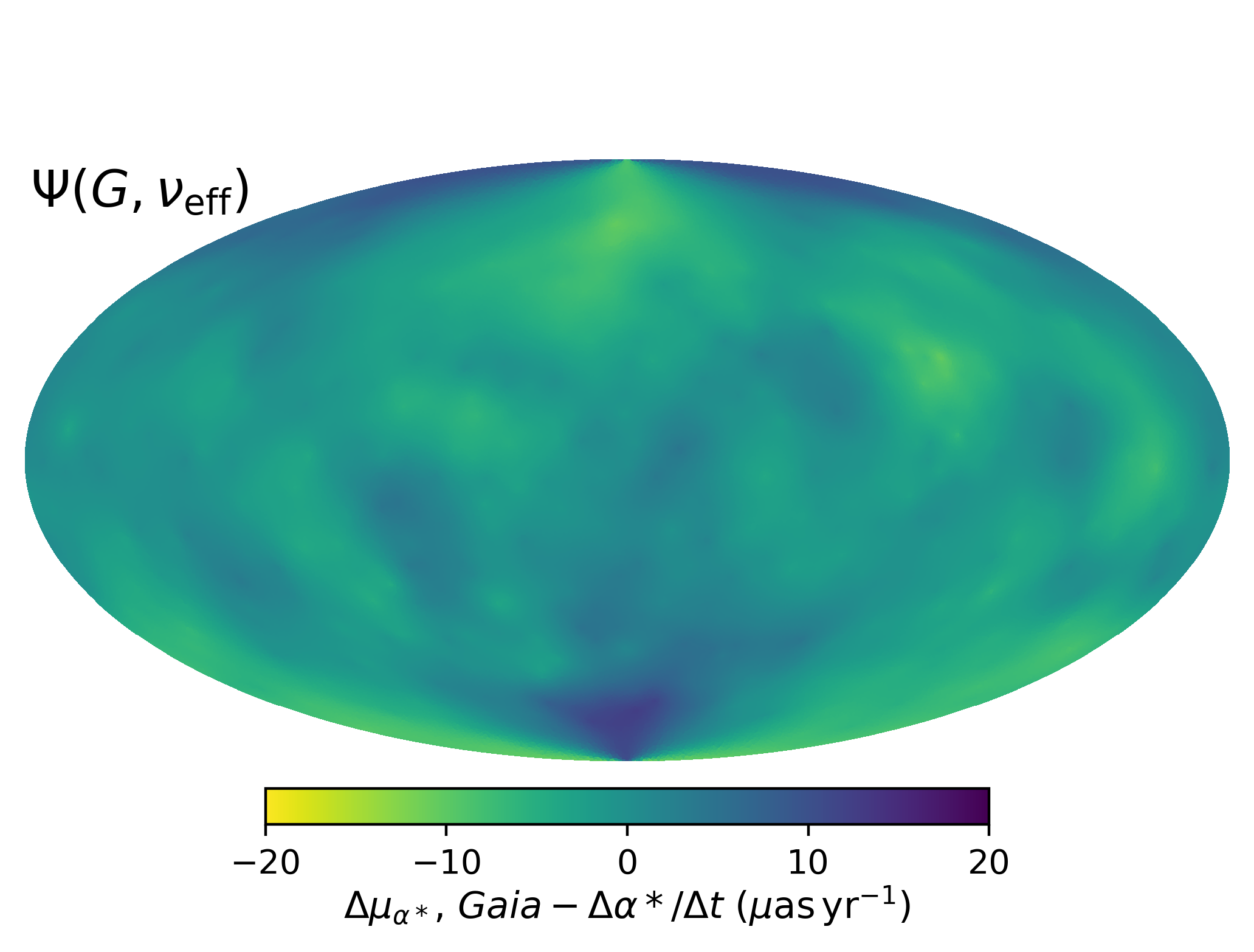}
\includegraphics[width=0.5\textwidth]{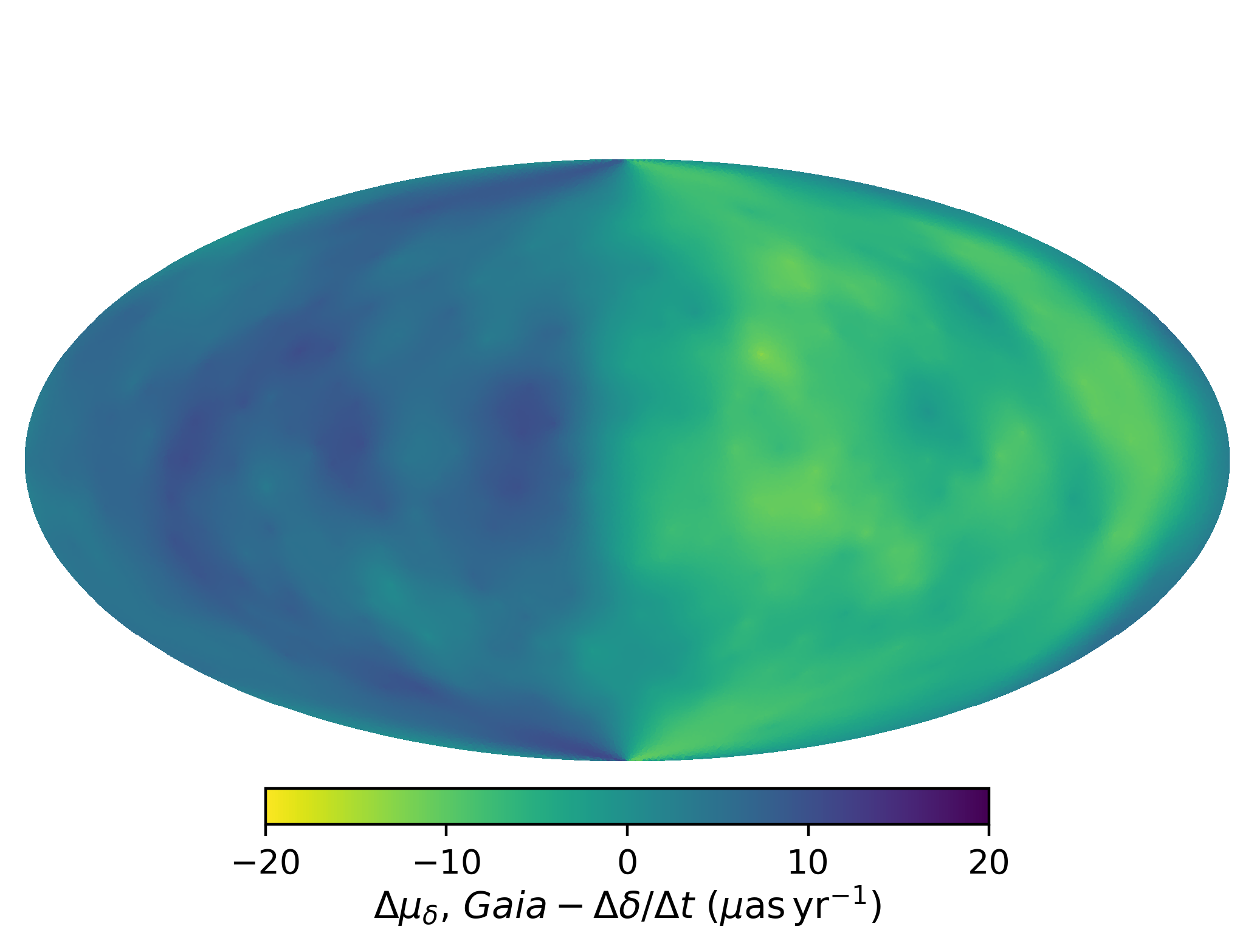}
\caption{Local corrections to the difference between \gaia EDR3 proper motions and long-term proper motions.  The top panels show the corrections independent of color and magnitude.  The lower panels show corrections $\Psi$ that depend on color and magnitude; Equations \eqref{eq:f_G_nueff} and \eqref{eq:pmcorrect} give the corrections for an individual star.  The maps are Hammer projections in equatorial coordinates with $\alpha = \delta = 0$ at the center; east is left and north is up.}
\label{fig:gaia_pmcorrect}
\end{figure*}

We follow the same approach for \gaia.  We use the same Gaussian mixture model as for DR2, but decrease the width of the broader, outlier Gaussian from 1~mas\,yr$^{-1}$ to 0.5~mas\,yr$^{-1}$ given \gaia's much improved precision.  \citetalias{Brandt_2018} used a coarser tiling of the sky than for \hipparcos proper motions, with more stars in each tile, to fit for the local frame rotation.  Here, we fit for frame rotations between \gaia EDR3 and the long-term proper motions in 2578 tiles, with an average of 45 stars per tile, compared to 4586 tiles (with 25 stars per tile) for \hipparcos.  Accounting for accelerating and cross-validation stars, we fit about 30 and 17 stars per tile for \gaia EDR3 and \hipparcos, respectively. 
The likelihood improvement for the cross-validation set of EDR3 proper motions is about $e^{290}$ over a constant frame rotation, and $e^{350}$ over no frame rotation; the corresponding improvements for the entire sample are about $e^{4900}$ and $e^{5400}$.  This again indicates only a modest degree of overfitting.  Increasing the number of tiles (decreasing the number of stars per tile) offers a negligible improvement in the cross-validation sample at the cost of substantial overfitting; decreasing the number of tiles by 40\% incurs a penalty of $\sim \!e^{20}$ in the likelihood of the cross-validation stars.  The top panels of Figure \ref{fig:gaia_pmcorrect} show the resulting corrections to the proper motions as a function of position in equatorial coordinates.

\cite{Fabricius+Luri+Arenou+etal_2020} find that the \gaia EDR3 reference frame does show some dependence on magnitude and color.  We test this possibility by using a coarser division of the sky and dividing our sample (again excluding the cross-validation set) by color and by magnitude.  We use 920 tiles for this step, with $\sim$100 stars per tile.  We split along the median color, $\nu_{\rm eff} \approx 1.57~\mu$m$^{-1}$, and at $G = 7.1$ (slightly below the median $G \approx 8$).  We choose $G=7.1$ as it turns out to provide the best agreement with the data, and it is close to the median $G$ magnitude.  It is also approximately the magnitude that separates the use of \gaia's lowest and its second-lowest gatings (gates being used to limit exposures and avoid saturation).  

We divide the sample sequentially.  We first split by color, separately fitting the redder half and the bluer half of stars.  We then split by magnitude and again separately fit the brighter and the fainter stars.  Our Gaussian process regression will then operate on the difference between the correction for red and blue stars, or between bright and faint stars.  

We next choose the functional forms that we assume for the dependence of an individual star's correction on $\nu_{\rm eff}$ and $G$ magnitude.
We assume that the correction will be proportional to $\nu_{\rm eff}$ relative to the catalog median normalized by its median absolute deviation, capping this value at $\pm 3$ based on our results in the cross-validation data set.  For magnitude, we instead choose a smoothed version of a Heaviside step function.  This is inspired by the shift in gating and based on improved results over a linear relationship.  We adopt the functional form
\begin{equation}
    f(G) = 1 - \frac{2}{1 + \exp \left(5(G - 7.1) \right)} .
\end{equation}
The factor of 5 in the exponent allows for a smooth transition over about a magnitude.  It performs slightly better in the cross-validation set than a step function and avoids a discontinuity.

Variations in the reference frame as a function of color and magnitude are shown in Figure \ref{fig:gaia_pmcorrect}, and are highly significant.  A color correction to the \gaia proper motions improves the likelihood of the cross-validation sample by a factor of $e^{168}$, equivalent to about $18\sigma$.  The improvement to the likelihood from the magnitude correction is lower at $e^{82}$, but still significant at about $13\sigma$.  Including both corrections improves the likelihood by $e^{257}$.  This is nearly equal to the product of the two improvements and indicates that the corrections may be regarded as independent.  Intriguingly, the correction maps for color and magnitude look almost identical.  Combining the two correction maps before smoothing with a Gaussian process regression slightly {\it improves} the likelihood of the cross-validation sample; it also reduces the potential for overfitting.  We therefore compute and use a single set of maps to correct for frame rotation as a function of both color and magnitude.  These maps are the ones that we use to compute likelihood ratios and that we show in Figure \ref{fig:gaia_pmcorrect}.  The actual correction applied to an individual star is the value in the map multiplied by a factor 
\begin{equation}
    f(G, \nu_{\rm eff}) = 1 - \frac{2}{1 + \exp \left(5(G - 7.1) \right)} + \frac{\nu_{\rm eff} - {\rm M}( \nu_{\rm eff})}{{\rm M}(|\nu_{\rm eff} - {\rm M}(\nu_{\rm eff})|)}
    \label{eq:f_G_nueff}
\end{equation}
where ${\rm M}(\nu_{\rm eff})$ denotes the median value of $\nu_{\rm eff}$; the denominator is the median absolute deviation.  The color and magnitude-dependent corrections for a given star are the product of Equation \eqref{eq:f_G_nueff} and the maps $\Psi(\alpha,\delta)$ shown in the lower panels of Figure \ref{fig:gaia_pmcorrect}.  The total cross-calibration term between the \gaia EDR3 and long-term proper motions is then
\begin{equation}
    \Delta \mu = {\rm Base} + f(G, \nu_{\rm eff}) \times \Psi.
    \label{eq:pmcorrect}
\end{equation}

For the full set of stars (about 10 times the cross-validation sample size), the likelihood improvements are nearly equal to the improvements in the cross-validation set raised to the tenth power.  For color, magnitude, and color+magnitude corrections, they are approximately $e^{1530}$, $e^{850}$, and $e^{2390}$, respectively.  These indicate an almost complete lack of overfitting and reflect the smoothness of the correction maps shown in Figure \ref{fig:gaia_pmcorrect}.

Most of the frame rotation that was present for the bright stars in \gaia DR2 was removed during the astrometric processing \citep{Lindegren+Klioner+Hernandez+etal_2020} and is not present in the maps shown in Figure \ref{fig:gaia_pmcorrect}.  Those rotations were easily the most apparent parts of their DR2 counterparts (Figure 6 of \citetalias{Brandt_2018}).  The maps still retain considerable structure, though the typical magnitude of this structure is a factor of $\sim$20 smaller than for the \hipparcos proper motions.  Both \hipparcos (when dividing the positions by 25 years) and \gaia EDR3 have similar uncertainties, and both contribute to the structure seen.  The magnitude and color-dependent effects are likely attributable to the \gaia proper motions; \cite{Fabricius+Luri+Arenou+etal_2020} reported color- and magnitude-dependent frame rotations of up to 0.1~mas\,yr$^{-1}$.  The structure seen in the top panels of Figure \ref{fig:gaia_pmcorrect} probably has a component attributable to the \hipparcos positions and one attributable to the \gaia EDR3 proper motions.

\section{The Gaia Uncertainties} \label{sec:gaia_uncertainties}

\citetalias{Brandt_2018} calibrated the \gaia DR2 uncertainties using the same Gaussian mixture model they used to calibrate the proper motions.  This succeeded for two reasons.  First, the long-term proper motions were considerably more precise than the DR2 proper motions; the \hipparcos positional uncertainties were relatively unimportant.  Second, the stars with real astrometric accelerations contributed little to the core of the distribution of residuals.  In other words, the core of the distribution shown in Figure 9 of \citetalias{Brandt_2018} appeared to be relatively uncontaminated by the distribution of real accelerators.  

\citetalias{Brandt_2018} found strong evidence that the ratio of actual uncertainties to formal uncertainties in \gaia DR2---the error inflation factor---was a function of position on the sky.  The analysis leading to this was possible only because of the low contamination of significant astrometric accelerators in the calibration sample.  The same analysis on \gaia EDR3 finds no evidence of spatial variation in the error inflation factor.  We speculate that the origin of this factor in DR2 may be associated with the DOF bug, which required an ad-hoc scaling of uncertainties.  This ad-hoc scaling was a function of magnitude but not of the number of transits used, or of ecliptic latitude.  It was through ecliptic latitude that the dependence was most visible in DR2 (see Figure 7 of \citetalias{Brandt_2018}).  

Our Gaussian mixture model with a constant variance for the background distribution turns out to be ill-suited to inferring the error inflation factor directly from the bulk of the data.
Figure \ref{fig:pm_residuals_raw} shows that real accelerators comprise a non-negligible fraction of stars even in the core of the distribution, with $z$-scores near zero.  In the following section we construct a calibration data set with almost no intrinsic astrometric acceleration suitable for calibrating the \gaia uncertainties.  Throughout this section, we neglect the fact that a few percent of the stars have six-parameter astrometric solutions (where the color is fit together with the position, parallax, and proper motion).  We have far too few stars to characterize any possible difference in the uncertainty calibration for these sources relative to the bulk of the stars with five-parameter fits.

\subsection{A Set of Reference Stars} \label{sec:rv_references}

To calibrate the uncertainties on the long-term proper motions and the \gaia EDR3 proper motions, we seek a sample of stars that are not accelerating within the precision of the measurements.  For this purpose we turn to long-term radial velocity monitoring with precision spectrographs.  We use the radial velocity time series as published by HIRES \citep{Butler+Vogt+Laughlin+etal_2017} and recalibrated by \cite{Tal-Or+Trifonov+Zucker+etal_2019}; the uniformly calibrated HARPS database \citep{Trifonov+Zechmeister+Kaminski+etal_2020}; and the older but long-baseline Lick sample \citep{Fischer+Marcy+Spronck_2014}.  

We take all \hipparcos targets in each of our three radial velocity data sets and fit a linear trend, adding a jitter to enforce a reduced $\chi^2$ of unity.  This subsumes any real companions, like planets, into a jitter term and/or a trend.  This is not a problem because we are searching for stars with low jitter and no trend.  Once we have determined the jitter, a simple linear fit returns the standard error of the trend, the radial velocity acceleration.  For stars that were observed by more than one spectrograph, we combine the measurements by each spectrograph using inverse-variance weighted means.

We next convert our radial velocity accelerations and their uncertainties to angular-equivalent units by multiplying by the \gaia EDR3 parallaxes.  We then convert these values into proper motion units by multiplying by 12.375 years.  This is half the time baseline between the \hipparcos and \gaia EDR3 catalog epochs, i.e., the effective time between the long-term proper motion and the EDR3 proper motion.  The resulting precision is directly comparable to the precision of the difference between the long-term proper motion and the \gaia EDR3 proper motion.  

We seek to define a sample of stars showing no astrometric acceleration.  Each star in our sample of candidate proper motion standard stars must meet five criteria:
\begin{enumerate}
    \item It must not have a common parallax companion in \gaia EDR3 within 10,000~AU in projection;
    \item Its radial velocity baseline must be at least five years on a single spectrograph;
    \item Its radial velocity jitter must be $<$10~m\,s$^{-1}$ for HIRES and HARPS, or $<$15~m\,s$^{-1}$ for Lick, to identify well-measured stars without planetary companions;
    \item Its radial velocity trend must be consistent with zero at $1\sigma$; and
    \item Its radial velocity trend precision must be better than the precision of the proper motion difference.
\end{enumerate}
For the last step, we take the \gaia EDR3 formal uncertainties and add these in quadrature with the formal uncertainties of the \hipparcos positions divided by 24.75 years.  We combine the published variances of the two \hipparcos reductions using a 60/40 mixture but apply no further error inflation.  We then average the proper motion precisions in right ascension and declination to get a single value for each star.

\begin{figure}
    \centering
    \includegraphics[width=\linewidth]{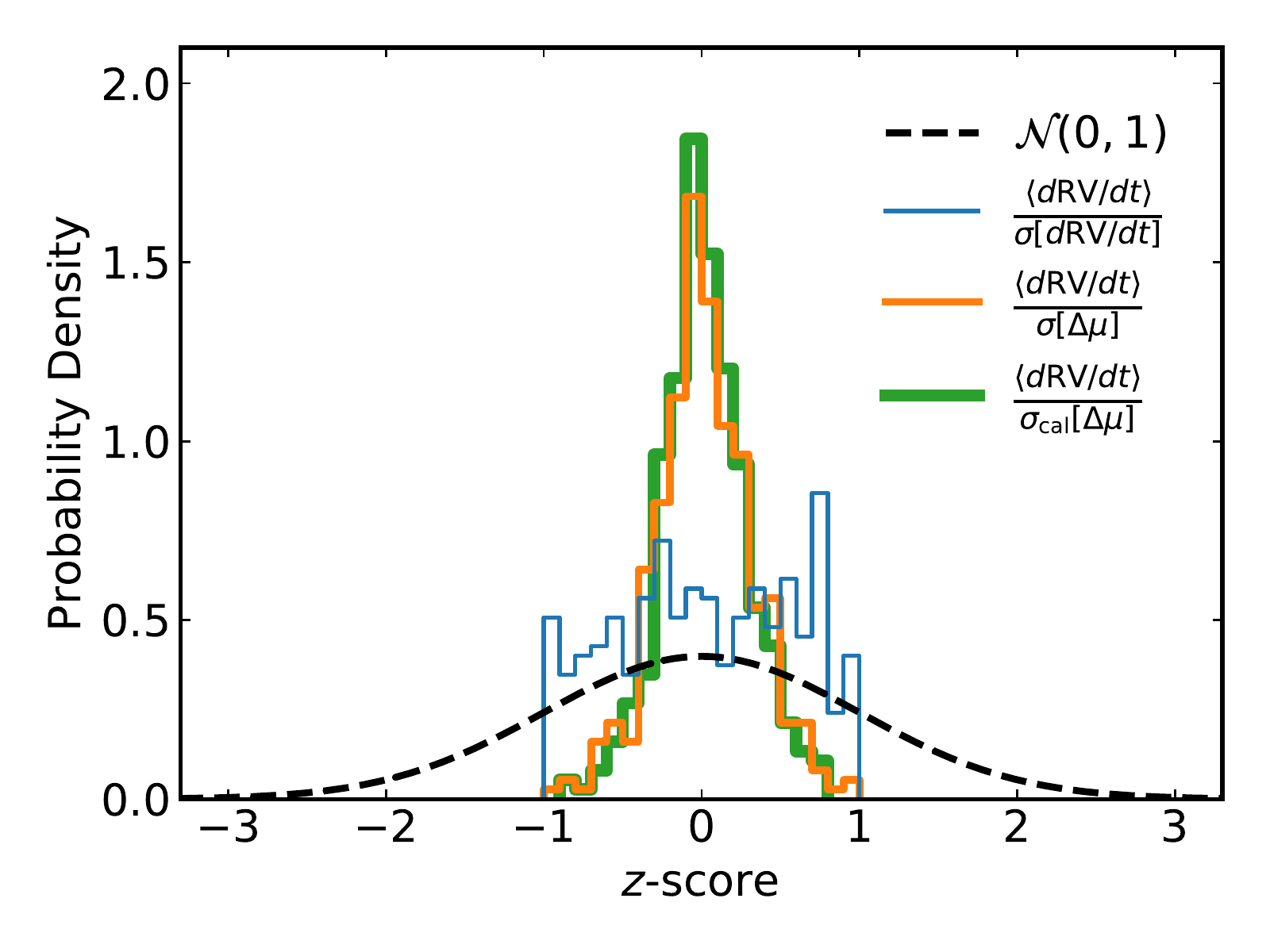}
    \caption{$z$-scores for the set of radial velocity benchmark stars we use to calibrate \gaia EDR3 uncertainties.  The thin blue line shows the best-fit trend divided by its standard error.  The thicker orange line normalizes the best-fit trend by the combined uncertainty from the \gaia EDR3 and long-term proper motions, divided by 12.375 years and converted to physical units.  The thickest green line adds the error inflation for \gaia EDR3 proper motions that we derive in Section \ref{sec:gaia_uncertainties}.  We use the average of the uncertainties in right ascension and declination for each star.  }
    \label{fig:RV_zscores}
\end{figure}

Our procedure results in a set of \nRVstandards astrometric standard stars.  Figure \ref{fig:RV_zscores} shows their $z$-scores, the ratio of the best-fit radial velocity trend normalized to three different uncertainties.  First, we normalize to the standard error of the radial velocity trend (after using jitter to enforce a reduced $\chi^2$ of unity).  Second, we normalize to the formal uncertainty in the proper motion difference after converting units and averaging the precisions in right ascension and declination.  Finally, we normalize to the calibrated uncertainties.  These latter two distributions, shown with progressively thicker lines in Figure \ref{fig:RV_zscores}, are much narrower than unit Gaussians: the limits on line-of-sight accelerations are significantly smaller than \gaia EDR3's precision for tangential accelerations.  Given our crude approach to modeling radial velocity jitter, we do not expect the distribution of $z$-scores to be accurately Gaussian, and we make no inference on the true distribution of radial velocity accelerations using this data set.

Some of our radial velocity standard stars could have low levels of real astrometric acceleration, e.g., from widely separated planets or brown dwarfs, but the purity of this sample will be far higher than for the catalog at large.  In the next section we will derive an error inflation by applying a slightly revised Gaussian mixture model to the radial velocity standards.

\subsection{Calibration}

We use the Gaussian mixture model given in Equation \eqref{eq:gaussmix} to fit for two error inflation terms: an additive term (in quadrature) with the \hipparcos positional uncertainties, and a multiplicative term for the \gaia EDR3 uncertainties:
\begin{align}
    \sigma^2 = a^2 \sigma^2[\mu_{G}] + \frac{0.6 \sigma^2[x_{H2}] + 0.4 \sigma^2[x_{H1}] + a^2 \sigma^2[x_{G}] + b^2}{(t_G - t_H)^2},
    \label{eq:error_inflation}
\end{align}
using $x$ to denote position in either right ascension or declination and the subscript to denote the catalog ($H1$\,=\,\cite{ESA_1997}, $H2$\,=\,\cite{vanLeeuwen_2007}, $G$\,=\,\gaia EDR3).  The additive term could be identified with \hipparcos positions and/or with \gaia EDR3 proper motions, while the multiplicative term is clearly associated with \gaia EDR3.  The covariance terms between right ascension and declination retain the factor $a$ but omit the term $b$.  A possible multiplicative error inflation on \hipparcos was ruled out by \citetalias{Brandt_2018}, and we see no evidence of one in the residuals that we derive below.  

We make two modifications to our fitting approach for Equation \eqref{eq:gaussmix}.  First, we take our outlier distribution to be a Gaussian with a covariance matrix 4 times that given by Equation \eqref{eq:error_inflation} (equivalent to 2 times the standard deviation) rather than one with a static width.  This better accommodates stars with widely varying precision.  Second, we fit for the probability $g$ that a star is not an outlier, i.e., that it is a good astrometric reference.  We also tried this alternative outlier distribution in the analysis of Section \ref{sec:refframe}, but found that it performed slightly worse under those circumstances.

\begin{figure*}
\begin{center}
    \includegraphics[width=0.5\linewidth]{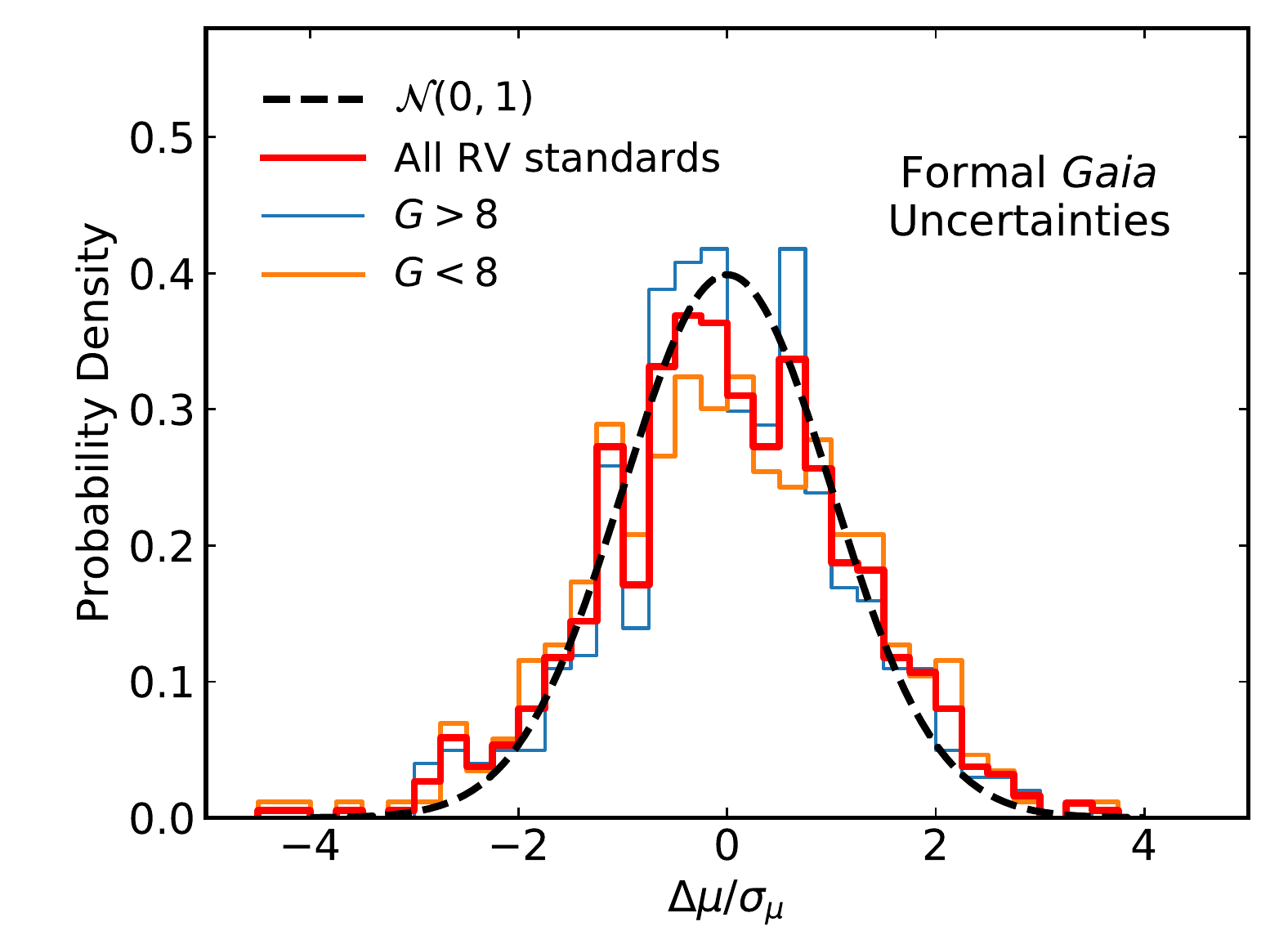}
    \hspace{-0.7 truein}
    \includegraphics[width=0.5\linewidth]{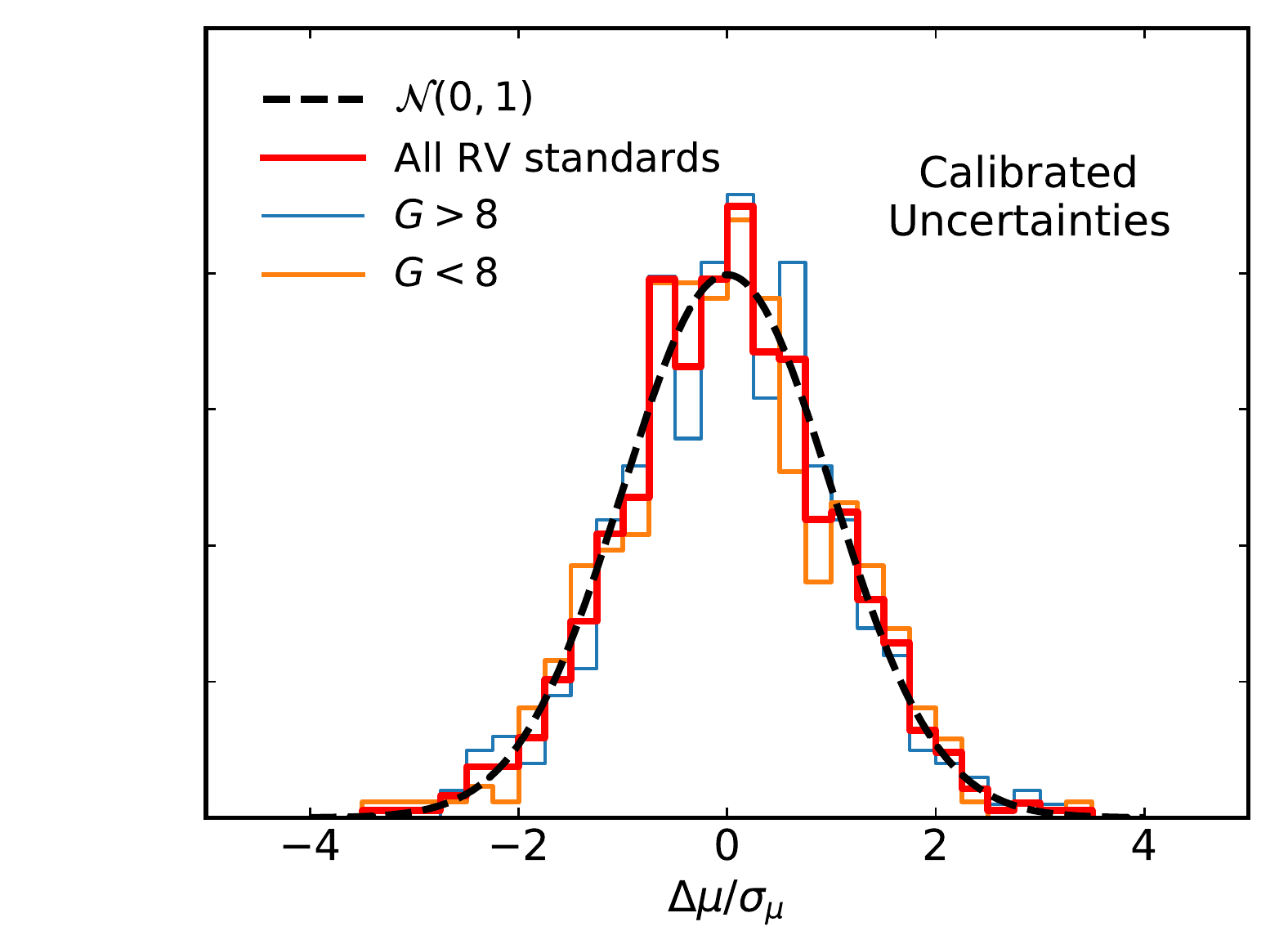} \\ \vskip -0.2 truein
\end{center}
    \caption{Proper motion residuals for our sample of \nRVstandards radial velocity standards without (left) and with (right) applying calibrations to the \gaia EDR3 astrometry.  The histograms include proper motions in both right ascension and declination.  The calibrated proper motion residuals (right panel) follow a Gaussian with unit variance.  The deviations from the unit Gaussian are consistent with Poisson statistics.}
    \label{fig:rv_standards}
\end{figure*}

Our approach returns a best-fit $a = 1.37$, $b = 0$, and $g=1$: all of the stars may be fit acceptably well by Equation \eqref{eq:error_inflation} and do not require an outlier distribution.  Figure \ref{fig:rv_standards} shows the $z$-scores of the astrometry of the radial velocity standards before (left) and after (right) applying this calibration to the uncertainties: the final distribution is accurately Gaussian with unit variance.  Our result for $a$ is smaller than the error inflation found by \citetalias{Brandt_2018} for \gaia DR2, and confirms EDR3's better control of systematic uncertainties \citep{Fabricius+Luri+Arenou+etal_2020, Lindegren+Klioner+Hernandez+etal_2020}.  The value of $b=0$~mas may be compared to the value of 0.2~mas\,yr$^{-1}$ found by \citetalias{Brandt_2018} for proper motion uncertainties in \hipparcos.  It suggests that no error inflation is needed on the \hipparcos positions and may, in part, be due to the use of \hipparcos positions to establish \gaia EDR3's bright star reference frame \citep{Lindegren+Klioner+Hernandez+etal_2020}.

We perform several consistency checks on our results.  Modest variations in the sample size and inclusion criteria have a minor effect on our inferred error inflation factor $a$.  We describe these variations, and the effects on our results, below.

First, we vary the width of the outlier distribution in Equation \eqref{eq:error_inflation}.  Changing its variance by a factor from two to sixteen (1.4 to 4 in standard deviation) does not affect the results.  As the outlier distribution becomes narrower still, it begins to merge with the central distribution and slightly decreases our inferred error inflation to $a \approx 1.30$
However, this brings a negligible benefit to the likelihood: about $e^{0.2}$ or $\Delta \ln {\cal L} = 0.2$.  As the outlier distribution merges with the central distribution the best-fit $g$ decreases; $g$ becomes effectively an additional free parameter.  The data themselves do not demand this extra flexibility.  For any relative width of the outlier distribution, we always infer an error inflation factor for \gaia EDR3 proper motions of at least 1.30.

We also perform consistency checks using bootstrap resampling on our set of radial velocity standard stars, and by varying the criteria for inclusion in this set.  
Bootstrap resampling gives a mean value of $a=1.35$ with a standard deviation of 0.08, and a mean value of $b=0.08$ with a standard deviation of 0.11.  The values of $a$ and $b$ are covariant: if we hold $b$ fixed at zero, the mean and standard deviation of $a$ become 1.37 and 0.07, respectively.  Loosening the criteria for inclusion in our radial velocity standard sample slightly increases our inferred error inflation and makes it more sensitive to the outlier distribution, likely due to the inclusion of astrometric accelerators.  
Increasing our stringency for radial velocity precision to require $0.7\sigma$ consistency with zero trend and a trend precision twice that of the astrometric precision yields just 142 reference stars, but this sample gives consistent best-fit values of $a=1.34$, $g=1$, and $b=0$.

Varying the sample of standard stars and the width of the outlier distribution change our inferred error inflation factor by only a few percent. 
We adopt $a=1.37$ at $b=0$ as our final values for the cross-calibration.  
Though we adopt $b=0$, it is difficult to distinguish a small value of $b$ from a small change in $a$ given our limited sample size: part of our inferred error inflation might be properly ascribed to the \hipparcos positions.  Some of the \hipparcos systematic uncertainties will also be shared by the \gaia EDR3 proper motions given the use of \hipparcos positions to establish the bright star reference frame \citep{Lindegren+Klioner+Hernandez+etal_2020}.  Any systematics shared by the two catalogs will cancel out in a cross-calibration.  If we adopt an additional uncertainty of 0.2~mas in \hipparcos (c.f.~the additional proper motion uncertainty inferred by \citetalias{Brandt_2018}), this would bring $a$ close to 1.3.  Adopting $a=1.30$ and $b=0.3$~mas rather than $a=1.37$ and $b=0$ produces distributions of residuals that are indistinguishable by eye from those shown in Figure \ref{fig:rv_standards}.
Continued radial velocity monitoring will ultimately produce better and larger sets of standard stars and enable a better characterization of \gaia's true astrometric uncertainties.

\section{Results} \label{sec:results}

\begin{figure*}
    \includegraphics[width=\linewidth]{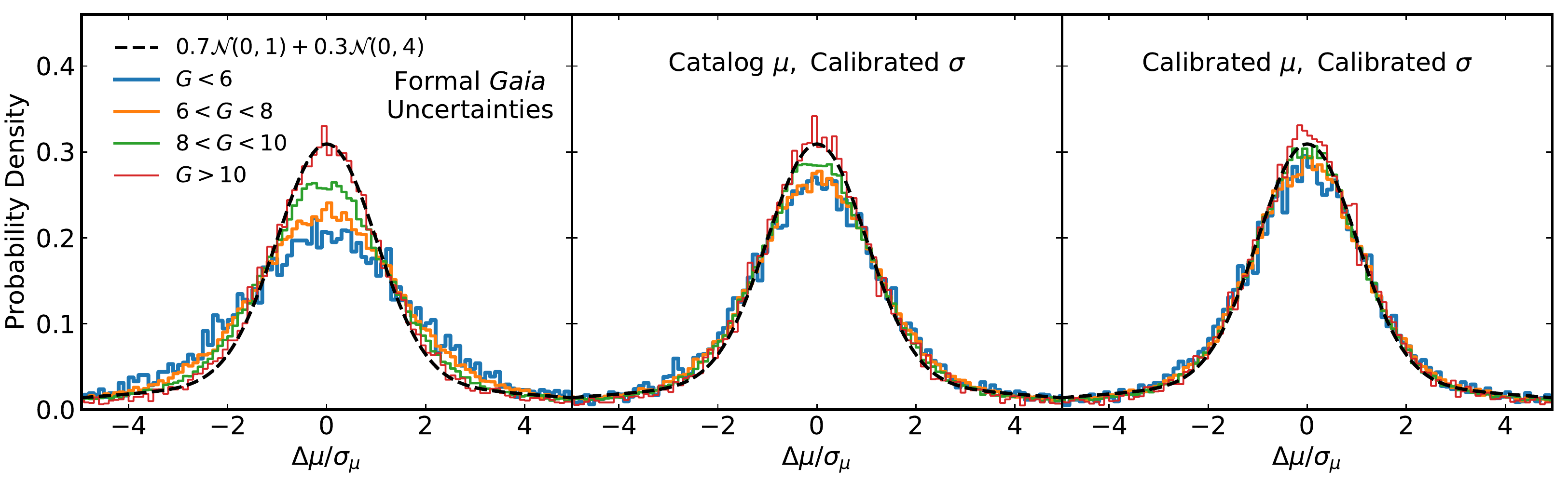}
    \caption{Differences between the \gaia EDR3 and long-term proper motions without (left) and with (right) calibrating the EDR3 proper motions and inflating their uncertainties.  The middle panel shows the calibrated uncertainties, but does not apply the local frame corrections shown in Figures \ref{fig:hip_pmcorrect} and \ref{fig:gaia_pmcorrect}; the left panel contains the same information as Figure \ref{fig:pm_residuals_raw}. The sum of Gaussians shown in black is a heuristic fit meant to guide the eye: about 70\% of stars shown appear to have constant proper motion at \gaia's current sensitivity.  An additional $\sim$25\% of stars have much larger proper motion residuals.  At fainter magnitudes ($G \gtrsim 10$), \hipparcos positional uncertainties dominate the error budget and calibrations to \gaia EDR3 proper motions have little effect.}
    \label{fig:residuals_final}
\end{figure*}

The results of the cross-calibration are three proper motions on a common reference frame with calibrated uncertainties.  For nearly all stars, the two most precise proper motions are the \gaia EDR3 proper motion and the long-term proper motion.  Figure \ref{fig:residuals_final} shows the residuals between these two proper motions, normalized by their uncertainties, before and after the calibrations we apply here.  We overplot the sum of two Gaussians as a reference to guide the eye; it is not intended to rigorously fit the data.  

The left panel of Figure \ref{fig:residuals_final} shows the increasing width of the distribution of $z$-scores (normalized residuals) with stellar brightness.  This reflects the fact that the formal \gaia EDR3 proper motion uncertainties underestimate the true uncertainties.  For faint stars, the \hipparcos positional uncertainties dominate the error budget.  The middle panel shows the effects of the calibrated \gaia proper motion uncertainties.  These bring the $z$-scores at different magnitudes into much better agreement with one another and with a unit Gaussian in the core.  The right panel shows the additional effect of the proper motion calibrations shown in Figure \ref{fig:gaia_pmcorrect}.  While the improvement does reflect some overfitting, our cross-validation sample shows that this overfitting is modest and that most of the improvement is real.  

Most of the stars shown in Figure \ref{fig:residuals_final} are not accelerating within the uncertainties of \gaia EDR3.  However, a substantial minority do appear to show astrometric acceleration.  The proper motion difference has two components (one each for right ascension and declination), so the relevant $\chi^2$ distribution has two degrees of freedom.  Overall, about 35,000 stars, or 30\% of the sample, have a $\chi^2$ value for the difference between their \gaia EDR3 and long-term proper motions of at least 11.8.  This threshold corresponds to a 0.3\% false alarm rate for Gaussian uncertainties.  Most of these stars have residuals too large for them to appear on Figure \ref{fig:residuals_final}.  Many more stars will show significant astrometric acceleration in future \gaia data releases.

\begin{figure*}
    \includegraphics[width=0.5\linewidth]{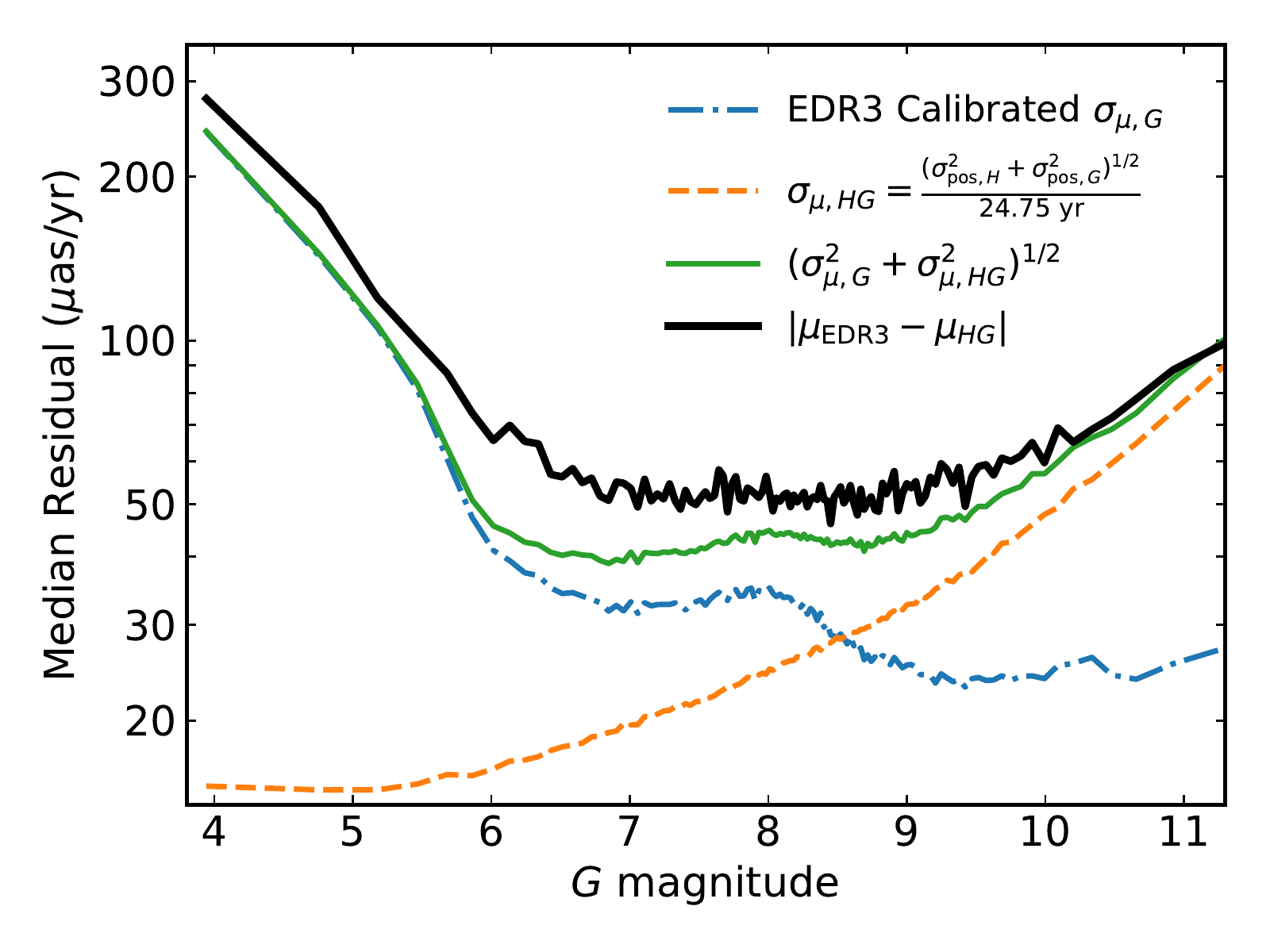}
    \includegraphics[width=0.5\linewidth]{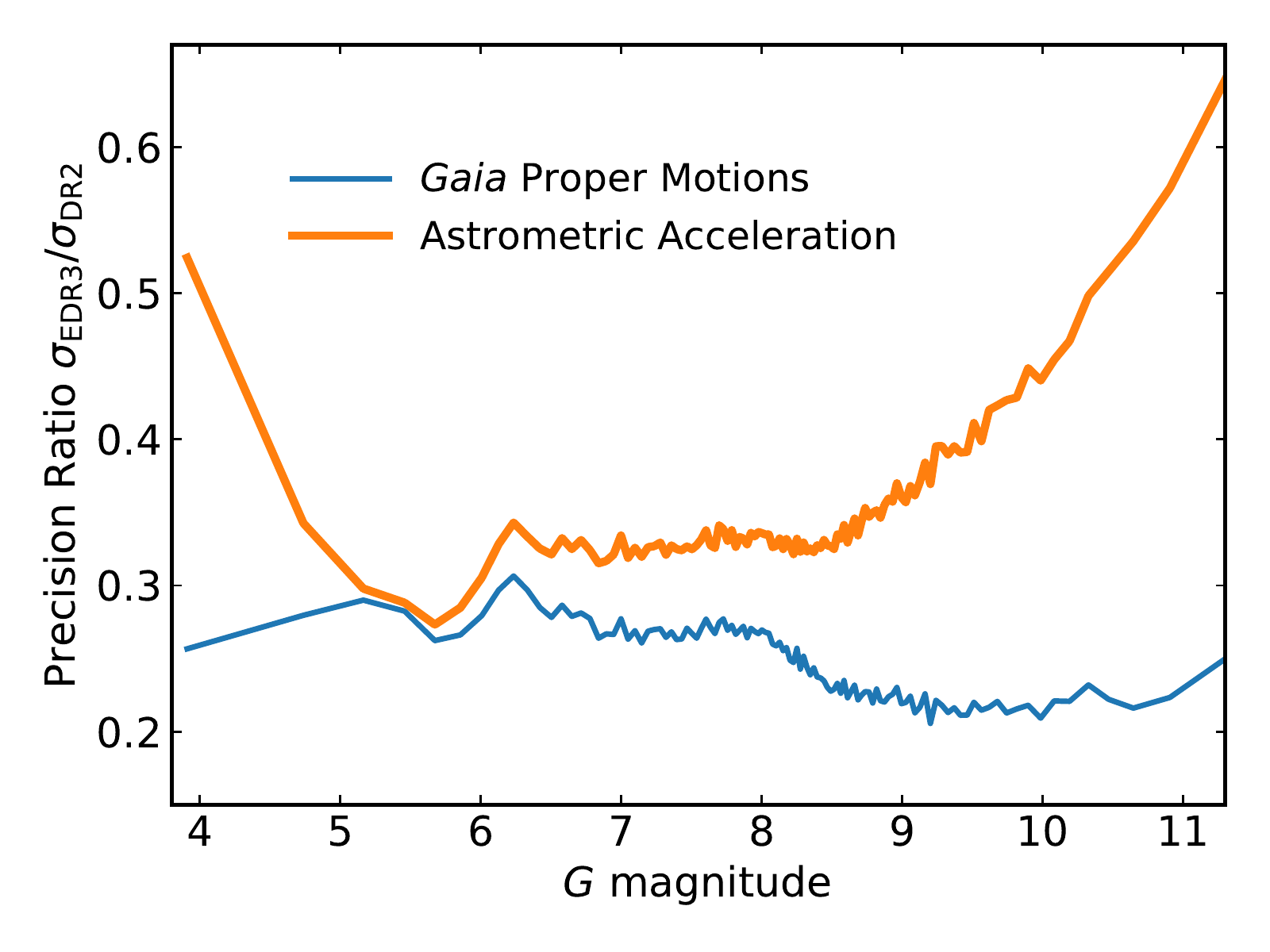}
    \caption{Left: median calibrated uncertainties as a function of magnitude, c.f.~Figure \ref{fig:errors_mag}.  Right: precision improvement of EDR3 over DR2 in the calibrated \gaia proper motions, and in the astrometric acceleration (the difference between the long-term proper motions and the more precise of the \gaia or \hipparcos proper motions).  The calibrated \gaia EDR3 proper motions represent a factor of $\sim$4 improvement over their DR2 counterparts.  This improvement is probably underestimated for very bright stars ($G \lesssim 6$); these showed evidence that their proper motion uncertainties remained underestimated even after error inflation (Figure 9 of \citetalias{Brandt_2018}).  The required error inflation for \gaia proper motions has fallen from a spatially dependent value averaging 1.7 in DR2 to a spatially uniform value of 1.37 in EDR3.  The EDR3 proper motions are now more precise than the long term proper motions for about half of all \hipparcos stars, and are more precise than \hipparcos proper motions even for very bright stars.  Overall, EDR3 offers a factor of $\sim$3 improvement in sensitivity to astrometric acceleration compared to DR2.  This factor is lower for $G \lesssim 5$, where \hipparcos proper motions were more precise than DR2 proper motions, and at faint magnitudes where \hipparcos positional uncertainties dominate the astrometric acceleration error budget.}
    \label{fig:uncertainties_final}
\end{figure*}

Figure \ref{fig:uncertainties_final} shows the final, calibrated uncertainties and residuals of the catalog as a function of magnitude, and the improvement over the DR2 edition of the HGCA.  The calibrated proper motion uncertainties of \gaia EDR3 are now more precise than the long-term proper motions for about 40\% of \hipparcos stars.  The typical precision of the proper motion difference is about 50~$\mu$as\,yr$^{-1}$ between magnitudes 6 and 9.  At a fiducial distance of 50\,pc and taking the timescale of this discrepancy to be 12.375 years, this translates to an acceleration precision of about 1~m\,s$^{-1}$\,yr$^{-1}$.  The precision is even better for closer stars.  It represents an improvement by a factor of $\sim$3 over the precision offered by the \gaia DR2 HGCA.  

\subsection{Use and Caveats}

The cross-calibration we provide assumes that the model sky paths used by \hipparcos and \gaia EDR3 are good approximations to the actual motions of the stars.  This is appropriate for single stars and for stars with much fainter companions on long-period orbits.  For these stars the acceleration between \hipparcos and \gaia will be large compared to the acceleration {\it within} either \hipparcos or \gaia.  Our cross-calibrated proper motions will be unreliable for stars with companions of comparable brightness.  In this case, \hipparcos, \gaia, or both will measure something closer to the photocenter motion of the system.  \hipparcos provides seven and nine-parameter fits to stars that showed astrometric acceleration within \hipparcos.  The \gaia EDR3 proper motions for these systems will likely be unreliable until the treatment of non-single stars in DR3.  

The cross-calibration should be taken with caution for extremely high proper motion stars.  As an example, Barnard's Star has a $\chi^2$ between the \gaia EDR3 and long-term proper motions of about 80, corresponding to more than $8\sigma$.  However, its discrepancy in proper motion in declination (0.56~mas\,yr$^{-1}$) is less than 4\% of the 16~mas\,yr$^{-1}$ correction for nonlinear motion.  The proper motion residual thus represents the apparent acceleration over about 6 months (4\% of 12.375 years).  Uncertainty in the true central epoch at which the proper motion is measured could account for much of this difference.  When astrometric accelerations are measured at very high significance, uncertainties in the effective epoch at which proper motions are measured can be an important contribution to uncertainties in orbit fitting.  This situation will be resolved when future \gaia data releases include the epoch astrometry.

Finally, we caution the user against making inferences using the distribution of $\chi^2$ residuals.  The catalog is intended to identify astrometrically accelerating stars for follow-up and to fit orbits to these systems.  The residuals will reflect a mixture of astrometric acceleration due to faint companions, photocenter motion for stars with bright companions, short-period orbits for which the \gaia sky paths are not yet good approximations, and possible residual systematics in the catalog.  However, for stars with faint, long-period companions, the combination of \hipparcos and \gaia provides a powerful tool for identifying systems, constraining orbits, and measuring masses.

\section{Construction and Structure of the Catalog} \label{sec:construction_structure}

The catalog contains three proper motions, three covariance matrices, the epochs at which these are measured, and cross-calibration terms.  
The structure of the catalog is identical to that of the HGCA \citepalias{Brandt_2018}, with one additional entry.  We provide the $\chi^2$ value (denoted {\tt chisq} in the table) between the two most precise proper motion measurements.  These are almost always \gaia EDR3 and the long-term proper motions.  
The $\chi^2$ values are not intended for population-level analyses but to select likely accelerators for follow-up by radial velocity or direct imaging surveys.  The computed values account for covariance between right ascension and declination, and have two degrees of freedom.  A false positive rate of 0.3\%, equivalent to three Gaussian sigma, translates to $\chi^2 \approx 11.8$.

The catalog provides the calibration terms between the various proper motions.  This includes cross-calibration offsets between the \hipparcos, \gaia EDR3, and the long-term proper motion reference frames.  It also includes the corrections for nonlinear apparent motion on the celestial sphere.  These corrections can be undone to recover the \hipparcos or the long-term proper motion in its own reference frame.  They are identical in name and form to their equivalents in the DR2 edition of the HGCA \citepalias{Brandt_2018}.  The \hipparcos proper motions in the \gaia EDR3 reference frame are given by, e.g.,
\begin{equation}
    \mu_{\alpha*,H} = f \mu_{\alpha*,{H2}} + (1 - f)\mu_{\alpha*,{H1}} + \xi_{\alpha*,H} + 2 \gamma_{\alpha*}
    \label{eq:crosscal}
\end{equation}
where $\mu_{H1}$ and $\mu_{H2}$ refer to the \cite{ESA_1997} and \cite{vanLeeuwen_2007} reductions of \hipparcos, respectively, and $f = 0.6$ is the weight given to the \hipparcos re-reduction (Section \ref{sec:hipparcos}).  The term $2\gamma$ is the correction for projecting linear motion onto the celestial sphere, and $\xi$ is the cross-calibration term between \hipparcos and \gaia EDR3.  We compute $\xi$ in practice as the sum of the calibration terms between \hipparcos and the long-term proper motion, and between the long-term proper motion and \gaia EDR3 proper motion.  The long-term proper motions in the \gaia EDR3 frame are given by 
\begin{equation}
    \mu_{\alpha*,HG} = \frac{f \alpha_{H2} + (1 - f)\alpha_{H1} - \alpha_G}{t_G - (f t_{H2} + (1 - f)t_{H1})}\cos \delta + \xi_{\alpha*,HG} + \gamma_{\alpha*}
    \label{eq:crosscal_hg}
\end{equation}
where $\alpha*$ refers to right ascension times cosine declination, $t$ refers to the astrometric epoch in a given catalog, and $\xi$, $\gamma$, and all subscripts have the same meanings as in Equation \eqref{eq:crosscal}.

\begin{deluxetable*}{lcrrrrrrrrrrr}
\tablewidth{0pt}
\tablecaption{The {\it Hipparcos}--{\it Gaia} Catalog of Accelerations: {\it Hipparcos} Proper Motions \label{tab:hgca_1}}
\tablehead{
    \colhead{{\it Hipparcos}} &
    \colhead{{\it Gaia} EDR3} &
    \colhead{$\mu_{\alpha*,H}$\tablenotemark{a}} &
    \colhead{$\sigma_{\alpha*,H}$} &
    \colhead{$\mu_{\delta,H}$\tablenotemark{a}} &
    \colhead{$\sigma_{\delta,H}$} &
    \colhead{Corr} &
    \colhead{$t_{\alpha*,H}$} &
    \colhead{$t_{\delta,H}$} &
    \colhead{$\xi_{\alpha*,H}$} &
    \colhead{$\xi_\delta,H$} &
    \colhead{$2\gamma_{\alpha*}$} &
    \colhead{$2\gamma_{\delta}$} 
    \\
    \colhead{Number} &
    \colhead{Source ID} &
    \multicolumn{2}{c}{mas\,yr$^{-1}$} &
    \multicolumn{2}{c}{mas\,yr$^{-1}$} &
    \colhead{} &
    \multicolumn{2}{c}{year} &
    \multicolumn{2}{c}{mas\,yr$^{-1}$} &
    \multicolumn{2}{c}{mas\,yr$^{-1}$}
    }
\startdata
 1 & 2738327528519591936 &    $-$4.87 &  1.31 &    $-$1.02 &  0.80 &    0.35 &  1991.55 &  1991.28 &     $-$0.03 &       0.47 &       0.00 &       0.00 \\
 2 & 2341871673090078592 &     183.28 &  1.41 &    $-$1.27 &  0.78 &    0.14 &  1991.48 &  1991.42 &    $-$0.38 &       0.04 &       0.00 &       0.00 \\
 3 & 2881742980523997824 &       4.65 &  0.49 &    $-$2.97 &  0.41 &    0.17 &  1990.85 &  1991.05 &       0.02 &       0.25 &       0.00 &       0.00 \\
 4 & 4973386040722654336 &      62.85 &  0.60 &       0.55 &  0.57 & $-$0.16 &  1991.01 &  1991.18 &       0.14 &       0.32 &       0.00 &       0.00 \\
 5 & 2305974989264598272 &       1.87 &  0.63 &       8.64 &  0.70 &    0.09 &  1991.10 &  1991.48 &    $-$0.15 &       0.07 &       0.00 &       0.00 \\
 6 & 2740326852975975040 &     224.05 &  5.77 &   $-$14.07 &  3.19 &    0.25 &  1991.34 &  1991.26 &     $-$0.02 &       0.29 &       0.00 &       0.00 \\
 7 & 2846308881856186240 &  $-$207.91 &  1.09 &  $-$200.99 &  0.78 &    0.41 &  1991.29 &  1991.23 &       0.08 &       0.21 &       0.00 &       0.00 \\
 8 & 2853169937491828608 &      19.04 &  1.32 &    $-$6.35 &  0.77 &    0.05 &  1991.57 &  1991.45 &     $-$0.06 &       0.19 &       0.00 &       0.00 \\
 9 & 2880160886370458368 &    $-$6.82 &  1.05 &       8.72 &  0.64 &    0.11 &  1991.26 &  1991.20 &    $-$0.16 &       0.25 &       0.00 &       0.00 \\
10 & 4976500987226833024 &      42.09 &  0.96 &      40.81 &  0.80 & $-$0.11 &  1991.23 &  1991.41 &       0.12 &       0.29 &       0.00 &       0.00 
 \enddata
 \tablenotetext{a}{Values include all local and nonlinearity corrections (e.g.~$\xi$ and $\gamma$, see Equation \eqref{eq:crosscal}).}
\end{deluxetable*}

\begin{deluxetable*}{lcrrrrrrrrr}
\tablewidth{0pt}
\tablecaption{The {\it Hipparcos}--{\it Gaia} Catalog of Accelerations: {\it Gaia} EDR3--{\it Hipparcos} Scaled Position Differences\label{tab:hgca_2}}
\tablehead{
    \colhead{{\it Hipparcos}} &
    \colhead{{\it Gaia} EDR3} &
    \colhead{$\mu_{\alpha*,HG}$\tablenotemark{a}} &
    \colhead{$\sigma_{\alpha*,HG}$} &
    \colhead{$\mu_{\delta,HG}$\tablenotemark{a}} &
    \colhead{$\sigma_{\delta,HG}$} &
    \colhead{Corr} &
    \colhead{$\xi_{\alpha*,HG}$} &
    \colhead{$\xi_{\delta,HG}$} &
    \colhead{$\gamma_{\alpha*}$} &
    \colhead{$\gamma_{\delta}$} 
    \\
    \colhead{Number} &
    \colhead{Source ID} &
    \multicolumn{2}{c}{mas\,yr$^{-1}$} &
    \multicolumn{2}{c}{mas\,yr$^{-1}$} &
    \colhead{} &
    \multicolumn{2}{c}{mas\,yr$^{-1}$} &
    \multicolumn{2}{c}{mas\,yr$^{-1}$} 
    }
\startdata
 1 & 2738327528519591936 &   $-$5.832 & 0.051 &   $-$5.093 & 0.029 &    0.37 &    $-$0.001 &      $-$0.004 &      0.000 &      0.000 \\
 2 & 2341871673090078592 &    181.510 & 0.048 &   $-$0.440 & 0.029 &    0.12 &   $-$0.012 &      0.008 &      0.000 &      0.001 \\
 3 & 2881742980523997824 &      5.762 & 0.014 &   $-$2.474 & 0.012 &    0.08 &      0.003 &      0.000 &      0.000 &      0.000 \\
 4 & 4973386040722654336 &     61.972 & 0.017 &      1.307 & 0.021 & $-$0.32 &      $-$0.023 &      0.014 &      0.000 &      0.000 \\
 5 & 2305974989264598272 &      0.986 & 0.022 &      8.744 & 0.022 &    0.07 &   $-$0.015 &      $-$0.012 &      0.000 &      0.000 \\
 6 & 2740326852975975040 &    223.167 & 0.174 &  $-$11.586 & 0.096 &    0.37 &    0.003 &      $-$0.015 &      0.000 &      0.000 \\
 7 & 2846308881856186240 & $-$211.040 & 0.039 & $-$197.017 & 0.031 &    0.33 &      0.038 &      $-$0.005 &      0.001 &      0.000 \\
 8 & 2853169937491828608 &     18.713 & 0.048 &   $-$6.635 & 0.031 &    0.02 &    0.008 &      $-$0.025 &      0.000 &      0.000 \\
 9 & 2880160886370458368 &   $-$6.043 & 0.034 &      9.244 & 0.021 &    0.01 &   0.000 &      $-$0.009 &      0.000 &      0.000 \\
10 & 4976500987226833024 &     42.321 & 0.028 &     40.833 & 0.028 & $-$0.11 &      $-$0.022 &      0.013 &      0.000 &      0.000 
 \enddata
 \tablenotetext{a}{Values include all local and nonlinearity corrections (e.g.~$\xi$ and $\gamma$, see Equation \eqref{eq:crosscal_hg}).}
\end{deluxetable*}

\begin{deluxetable*}{lcrrrrrrrrrr}
\tablewidth{0pt}
\tablecaption{The {\it Hipparcos}--{\it Gaia} Catalog of Accelerations: {\it Gaia} EDR3 Proper Motions\label{tab:hgca_3}}
\tablehead{
    \colhead{{\it Hipparcos}} &
    \colhead{{\it Gaia} EDR3} &
    \colhead{$\mu_{\alpha*,G}$\tablenotemark{a}} &
    \colhead{$\sigma_{\alpha*,G}$} &
    \colhead{$\mu_{\delta,G}$\tablenotemark{a}} &
    \colhead{$\sigma_{\delta,G}$} &
    \colhead{Corr\tablenotemark{a}} &
    \colhead{$t_{\alpha*,G}$} &
    \colhead{$t_{\delta,G}$} &
    \colhead{$\chi^2[\Delta \mu]$}
    \\
    \colhead{Number} &
    \colhead{Source ID} &
    \multicolumn{2}{c}{mas\,yr$^{-1}$} &
    \multicolumn{2}{c}{mas\,yr$^{-1}$} &
    \colhead{} &
    \multicolumn{2}{c}{year} &
    \colhead{($N_{\rm d.o.f.} = 2$)}
    }
\startdata
 1 & 2738327528519591936 &   $-$0.360 & 0.079 &   $-$5.053 & 0.030 &    0.11 &  2015.93 &  2015.53 &   3515.5\\
 2 & 2341871673090078592 &    179.805 & 0.783 &   $-$1.041 & 0.579 &    0.07 &  2015.82 &  2015.51 &      5.5\\
 3 & 2881742980523997824 &      5.761 & 0.038 &   $-$2.406 & 0.032 &    0.19 &  2016.61 &  2016.14 &      4.2\\
 4 & 4973386040722654336 &     61.965 & 0.019 &      1.302 & 0.025 & $-$0.33 &  2015.97 &  2016.02 &      0.2\\
 5 & 2305974989264598272 &      1.022 & 0.028 &      8.733 & 0.026 &    0.03 &  2016.38 &  2015.88 &      1.1\\
 6 & 2740326852975975040 &    223.197 & 0.040 &  $-$11.498 & 0.026 &    0.23 &  2015.95 &  2015.43 &      0.8\\
 7 & 2846308881856186240 & $-$206.509 & 0.039 & $-$196.098 & 0.017 & $-$0.10 &  2015.62 &  2015.34 &   6859.7\\
 8 & 2853169937491828608 &     18.746 & 0.094 &   $-$6.472 & 0.049 & $-$0.22 &  2016.07 &  2015.82 &      8.6\\
 9 & 2880160886370458368 &   $-$6.071 & 0.028 &      9.296 & 0.021 & $-$0.15 &  2015.94 &  2015.85 &      3.3\\
10 & 4976500987226833024 &     42.298 & 0.014 &     40.841 & 0.018 & $-$0.38 &  2016.01 &  2016.10 &      0.6
\enddata
\tablenotetext{a}{Values are identical to those published in {\it Gaia} EDR3 \citep{Lindegren+Klioner+Hernandez+etal_2020}.}
\end{deluxetable*}

Tables \ref{tab:hgca_1}, \ref{tab:hgca_2}, and \ref{tab:hgca_3} list the first ten rows of the \gaia EDR3 edition of the HGCA.  Table \ref{tab:hgca_1} is very similar to Table 2 of \citetalias{Brandt_2018}. The proper motions are systematically offset from their values in the DR2 edition of the HGCA because they are calibrated to the EDR3 reference frame rather than the rotating DR2 reference frame.  Table \ref{tab:hgca_2} likewise closely matches Table 3 in \citetalias{Brandt_2018}.  Table \ref{tab:hgca_3} shows considerable improvements in sensitivity compared to the DR2 proper motions (Table 4 of \citetalias{Brandt_2018}).  Two of the first ten \hipparcos stars, HIP~1 and HIP~7, are accelerating at high significance.  The acceleration in right ascension for HIP~1, for example, is significant at nearly $60\sigma$.  The $\chi^2$ values in the last column show that the other eight stars are not accelerating at an equivalent significance of $3\sigma$ ($\chi^2 = 11.8$).  HIP~8 has $\chi^2 = 8.6$, which is about $2.5\sigma$ significant.

The catalog format and data fields are identical to the DR2 edition of the HGCA apart from the $\chi^2$ field that we have added for convenience.  Table \ref{tab:hgca_fields} lists the field names and their definitions.  It is almost identical to Table 5 of \citetalias{Brandt_2018}.

\section{Conclusions} \label{sec:conclusions}

In this paper we have derived a cross-calibration of \hipparcos and \gaia EDR3 suitable for identifying astrometrically accelerating systems and for fitting orbits.  The catalog contains just over 115,000 stars, of which about 30\% are inconsistent with constant proper motion at 99.7\% (equivalent to $3\sigma$ in a Gaussian distribution).  

Our approach to the cross-calibration matches that of \citetalias{Brandt_2018}, and our results for \hipparcos proper motions carry over largely unchanged.  This consists of a 60/40 mixture of the \cite{vanLeeuwen_2007} and \cite{ESA_1997} reductions with an additional uncertainty of $0.2$~mas\,yr$^{-1}$ added in quadrature with the proper motion uncertainties, and a locally variable frame rotation of up to $\sim$1~mas\,yr$^{-1}$.  We have confirmed the 60/40 mix of the two \hipparcos reductions using only \hipparcos positions.

The \gaia Collaboration removed the global frame rotation from EDR3 using the \hipparcos positional measurements.  As a result, the global frame rotation that was a dominant aspect of the cross-calibration for DR2 is gone.  However, we find strong evidence for a locally variable cross-calibration, and for locally-variable frame rotation as a function of color and magnitude.  The latter result is consistent with the findings of the \gaia Collaboration \citep{Fabricius+Luri+Arenou+etal_2020}.

The \gaia EDR3 proper motion uncertainties are more difficult to calibrate than their DR2 counterparts.  For this purpose, we turn to long-term radial velocity surveys to construct a sample of astrometric reference stars.  We find an error inflation factor, a ratio of total to formal uncertainties, of 1.37.  Varying the details of the analysis and of the sample of radial velocity standard stars results in inferred inflation factors ranging from about 1.30 to just over 1.40.  An error inflation of 1.37 is broadly consistent with the results for fainter stars \citep{Fabricius+Luri+Arenou+etal_2020} and is substantially lower than the spatially-variable factor of $\sim$1.7 found by \citetalias{Brandt_2018} for \gaia DR2.  This reflects the improvements in data processing in EDR3, while the lack of spatial dependence may be related to the fixing of the DOF bug \citep{Gaia_Astrometry_2018}.

The resulting catalog represents a factor of $\sim$3 improvement in sensitivity to astrometric accelerations over the \gaia DR2 edition of the HGCA.  The improvement varies with magnitude.  It is lower for faint stars where the uncertainty is already dominated by the \hipparcos positional uncertainties, and it is lower for very bright stars where the \hipparcos proper motions were more precise than the \gaia DR2 proper motions.  

\hipparcos positional measurements will continue to provide valuable data points for bright stars even in future \gaia data releases that treat astrometrically accelerating stars internally.  Orbital fits derived by our cross-calibration will also provide an independent check on the accelerations inferred by \gaia DR3.  \\

{\it Software}: astropy \citep{astropy:2013, astropy:2018},
          scipy \citep{2020SciPy-NMeth},
          numpy \citep{numpy1, numpy2}

\acknowledgments{T.D.B.~thanks Daniel Michalik and G.~Mirek Brandt for input on a draft of this paper, and an anonymous referee for corrections and suggestions for improvement. T.D.B.~gratefully acknowledges support from National Aeronautics and Space Administration (NASA) under grant 80NSSC18K0439 and from the Heising-Simons Foundation (HSF) under grant 20190295.  This work has made use of data from the European Space Agency (ESA) mission Gaia (https://www.cosmos.esa.int/gaia), processed by the Gaia Data Processing and Analysis Consortium (DPAC; https://www.cosmos.esa.int/web/gaia/dpac/consortium). Funding for the DPAC has been provided by national institutions, in particular the institutions participating in the Gaia Multilateral Agreement.}

\begin{deluxetable*}{lcr}
\tablewidth{0pt}
\tablecaption{The {\it Hipparcos}--{\it Gaia} Catalog of Accelerations, EDR3 Edition: Description of Catalog Contents\label{tab:hgca_fields}}
\tablehead{
    \colhead{Parameter Name} &
    \colhead{Units} &
    \colhead{Description}
    }
\startdata
${\tt hip\_id}$ & & {\it Hipparcos} identification number \\
${\tt gaia\_source\_id}$ & & {\it Gaia} EDR3 source identification number \\
${\tt gaia\_ra}$ & degrees & {\it Gaia} EDR3 measured right ascension \\
${\tt gaia\_dec}$ & degrees & {\it Gaia} EDR3 measured declination \\
${\tt radial\_velocity}$ & km\,s$^{-1}$ & Measured radial velocity \\
${\tt radial\_velocity\_error}$ & km\,s$^{-1}$ & Radial velocity standard error \\
${\tt radial\_velocity\_source}$ & & Source of measured radial velocity, \gaia DR2 or XHIP \citep{XHIP_2012} \\
${\tt parallax\_gaia}$ & mas & {\it Gaia} EDR3 parallax \\
${\tt parallax\_gaia\_error}$ & mas & {\it Gaia} EDR3 parallax standard error \\
${\tt pmra\_gaia}$ & mas\,yr$^{-1}$ & {\it Gaia} EDR3 proper motion in right ascension, $d\alpha/dt \cos \delta$ \\
${\tt pmra\_gaia\_error}$ & mas\,yr$^{-1}$ & Calibrated uncertainty in ${\tt pmra\_gaia}$ \\
${\tt pmdec\_gaia}$ & mas\,yr$^{-1}$ & {\it Gaia} EDR3 proper motion in declination \\
${\tt pmdec\_gaia\_error}$ & mas\,yr$^{-1}$ & Calibrated uncertainty in ${\tt pmdec\_gaia}$ \\
${\tt pmra\_pmdec\_gaia}$ & & Correlation between ${\tt pmra\_gaia}$ and ${\tt pmdec\_gaia}$ \\
${\tt pmra\_hg}$ & mas\,yr$^{-1}$ & Calibrated proper motion in right ascension from the {\it Hipparcos}--{\it Gaia} positional difference \\
${\tt pmra\_hg\_error}$ & mas\,yr$^{-1}$ & Calibrated uncertainty in ${\tt pmra\_hg}$ \\
${\tt pmdec\_hg}$ & mas\,yr$^{-1}$ & Calibrated proper motion in declination from the {\it Hipparcos}--{\it Gaia} positional difference \\
${\tt pmdec\_hg\_error}$ & mas\,yr$^{-1}$ & Calibrated uncertainty in  ${\tt pmdec\_hg}$ \\
${\tt pmra\_pmdec\_hg}$ & & Correlation between ${\tt pmra\_hg}$ and ${\tt pmdec\_hg}$ \\
${\tt pmra\_hip}$ & mas\,yr$^{-1}$ & Calibrated proper motion in right ascension from the composite {\it Hipparcos} catalog \\
${\tt pmra\_hip\_error}$ & mas\,yr$^{-1}$ & Calibrated uncertainty in ${\tt pmra\_hip}$ \\
${\tt pmdec\_hip}$ & mas\,yr$^{-1}$ & Calibrated proper motion in declination from the composite {\it Hipparcos} catalog \\
${\tt pmdec\_hip\_error}$ & mas\,yr$^{-1}$ & Calibrated uncertainty in  ${\tt pmdec\_hip}$ \\
${\tt pmra\_pmdec\_hip}$ & & Correlation between ${\tt pmra\_hip}$ and ${\tt pmdec\_hip}$ \\
${\tt epoch\_ra\_gaia}$ & year & Central epoch of {\it Gaia} EDR3 right ascension measurement \\
${\tt epoch\_dec\_gaia}$ & year & Central epoch of {\it Gaia} EDR3 declination measurement \\
${\tt epoch\_ra\_hip}$ & year & Central epoch of {\it Hipparcos} right ascension measurement \\
${\tt epoch\_dec\_hip}$ & year & Central epoch of {\it Hipparcos} declination measurement \\
${\tt crosscal\_pmra\_hg}$ & mas\,yr$^{-1}$ & Difference in ${\tt pmra\_hg}$ from the catalog-computed value: $\xi_{\alpha*,HG}$ from Table \ref{tab:hgca_2} \\
${\tt crosscal\_pmdec\_hg}$ & mas\,yr$^{-1}$ & Difference in ${\tt pmdec\_hg}$ from the catalog-computed value: $\xi_{\delta,HG}$ from Table \ref{tab:hgca_2} \\
${\tt crosscal\_pmra\_hip}$ & mas\,yr$^{-1}$ & Difference in ${\tt pmra\_hip}$ from the catalog-computed value: $\xi_{\alpha*,H}$ from Table \ref{tab:hgca_1} \\
${\tt crosscal\_pmdec\_hip}$ & mas\,yr$^{-1}$ & Difference in ${\tt pmra\_hip}$ from the catalog-computed value: $\xi_{\delta,H}$ from Table \ref{tab:hgca_1} \\
${\tt nonlinear\_dpmra}$ & mas\,yr$^{-1}$ & Correction to ${\tt pmra\_hg}$ from projecting linear motion onto the celestial sphere: $\gamma_{\alpha*}$ from Table \ref{tab:hgca_2} \\
${\tt nonlinear\_dpmdec}$ & mas\,yr$^{-1}$ & Correction to ${\tt pmdec\_hg}$ from projecting linear motion onto the celestial sphere: $\gamma_{\delta}$ from Table \ref{tab:hgca_2}  \\
${\tt chisq}$ & & $\chi^2$ value for a model of constant proper motion with 2 degrees of freedom
\enddata
\end{deluxetable*}

\bibliographystyle{apj_eprint}
\bibliography{refs.bib}

\end{document}